\documentclass[aps,pra,superscriptaddress,twocolumn,amsmath,amsfonts,amssymb,floatfix]{revtex4-1}
\usepackage{enumerate}
\usepackage{mathtools}
\usepackage{amsmath}
\usepackage{graphicx}
\usepackage{epsfig}
\usepackage{sidecap}
\usepackage{hyperref}
\usepackage{color}
\usepackage{subfigure,array}
\usepackage{verbatim}
\usepackage{multirow}
\usepackage{datetime}
\usepackage{xcolor}
\usepackage[T1]{fontenc}

\usepackage{enumerate}
\usepackage{bm}

\begin{document}

\title{Josephson Oscillation and Nonlinear Self-Trapping in Quasi-one-dimensional Quantum Liquid}

\author{Shivam Singh}
\author{Ibrar}
\author{Suhail Rashid}
\author{Ayan Khan} \thanks{ayan.khan@bennett.edu.in}
\affiliation{Department of Physics, School of Engineering and Applied Sciences, Bennett University, Greater Noida, UP-201310, India}
\begin{abstract}
In this article, we study the two-mode method to analyze the Josephson oscillation for a trapped binary Bose-Einstein condensate while taking into account the beyond mean-field and three body interactions. For this purpose, we use the archetypal model of double well potential and study the Josephson oscillation and self-trapping phases in quasi-one dimension. Additionally, our analysis provides quantitative discussion on the effect of asymmetry and dimension. We further corroborate our findings with Bogoliubov quasi-particle method and notice the existence of roton-like mode.
\end{abstract}

\date{\today}

\maketitle
\section{Introduction} 
Macroscopic Quantum Tunnelling (MQT) is a matter of deep interest due to its unique non-classical nature and potential applications in quantum computation and sensing \cite{devoret}. In recent times, MQT in superconducting circuits turned out a reality following the mechanism of Josephson junction \cite{nori,yu,martinis}. The coherent oscillations
between quantized levels in a current biased Josephson junction has lead to the generation of phase qubit and opened new avenues for quantum information processing.

However, MQT in Bose-Einstein condensates (BEC) is matter of considerable attention for last couple of decades \cite{cazalilla2011one,dutta2015tunneling}. BEC provide an unique platform for exploring quantum phenomena at macroscopic scale. The unprecedented tunability and controllability of the ultra-cold system allows us 
to investigate not only fundamental quantum phenomena but also promises to realize quantum devices leading to atomtronics, quantum metrology \cite{nature_Qmetro_2021BEC,amico} and quantum sensing \cite{degen_2017quantum,schafer}.

In the context of BEC, this involves the tunnelling of the macroscopic
condensate wave-function described by the Gross-Pitaevskii (GP) equation.
Unlike single-particle tunnelling, MQT is governed by nonlinear interactions,
coherence properties and collective energy landscapes. One of the most well studied system in this context is the double well potential.
In double-well potentials, MQT manifests as Josephson tunnelling of the
condensate population between wells \cite{smerzi}. The dynamics can be described using
a two-mode approximation, where the population imbalance and relative
phase serve as conjugate variables \cite{smerzi,gati,sofia}. 
Quantum tunnelling can occur between
meta-stable states, such as self-trapped imbalanced configurations. 
Very recently the idea has also been extended in the context of atom-phonon interaction and `self-trapped limit cycle' is being reported \cite{vivek}.
The rate
of tunnelling is determined by the curvature of the effective GP potential
landscape and exhibits dependence on interaction strength, barrier shape
and condensate density \cite{smerzi}.


The recently observed quantum droplets emerging from BEC due to the clustering of atoms via delicate balance between the mean-field (MF) and beyond mean-field (BMF) interactions have paved a way towards the uncharted territories of the quantum world further \cite{kadau,pfau,cab,che}. The emergence of liquid-like state is understood from the perspective of interaction competition in the mean-field level and contribution of higher order interaction pertaining to Lee-Huan-Yang (LHY) theory \cite{lee}. It is understood that when all the interactions in the mean-field level compensate each other, a small beyond mean-field contribution (LHY) supports the formation of the droplets \cite{petrov,petrov_1d,malomed1,deb}. These droplets are considered as an ideal testing ground for quantum many-body interactions \cite{ifb}.

The initial experimental verification of droplets were carried out on dipolar BEC \cite{kadau,pfau} and binary BEC \cite{cab,che}. Later it has been observed that insignificant LHY contribution can also lead to the formation of super-solid like phase \cite{tanzi2019observation, chomaz2019long, bottcher2019transient} in dipolar BEC. This unique state, which is a superfluid with lattice ordering, can be captured through the investigation of Bogolliubov dispersion where the emergence of roton-like mode is attributed to transition toward the super-solid like phase from Bose liquid \cite{chomaz, roccuzzo2019supersolid}. 


It is well understood that the fundamental attributes of a BEC is revealed through its response to perturbations, which is governed by its elementary excitations \cite{pethick2008bose, pitaevskii_2016bose, mukherjee2025collective}. These excitations are crucial to determine the thermodynamic and dynamic properties of the condensate \cite{pethick2008bose, pitaevskii_2016bose, leggett_1999superfluidity}, such as superfluidity, sound propagation, and collective modes \cite{pethick2008bose, pitaevskii_2016bose, pethick2008bose, pitaevskii_2016bose}. 

In the context of quantum liquid, arising from dipolar BEC and binary BEC, albeit their same stabilization mechanism, each one has its own characteristics.
Quantum liquids of dipolar atoms are in general anisotropic in nature, this is due to the fact that the attractive dipolar interaction tries to align the atoms in the dipole direction while the surface tension supports a round droplet. This leads to a scissor-like oscillation resulting an angular oscillation of the droplet about the dipole's axis and accordingly named as scissors mode \cite{ferrier2018scissors}. In two component BEC, the density
ratio between the two species is fixed by the mean-field interactions. Nevertheless, there is a possibility of overabundance of a single species of untrapped atoms making a halo around the droplet.
They exhibit excitations where the two components move either in or out of phase relative to each other. The collective oscillation mode is described as ripplons which is very similar to the normal liquid and arising from the surface tension \cite{petrov}. An accurate knowledge of their spectrum of collective excitation promises to provide valuable information about their precise equation of state.

Off late, elementary excitation and Josephson oscillation (JO) were studies for quantum liquid in one dimension \cite{P.wysocki_josephson_2024, abdullaev2024beyond}. The role of dimensionality in quantum fluctuation and vortex realization in Bosonic Josephson junction have also been reported recently \cite{bardin2024quantum, bellettini2024massive}. It is been pointed out that the Josephson oscillation can be used as a tool to study the superfluid fraction in a supersolid at a very low density regime \cite{biagioni2024measurement}. Very recently, an exciting new study has emerged in the context of qubit preparation where a single impurity atom trapped in a double well potential created by a dipolar two-soliton molecule in a quasi-one-dimensional geometry \cite{al2026single}.

It is noteworthy that, in one-dimension, the mean-field interaction is repulsive while the beyond mean-field contribution is attractive \cite{petrov_1d}. This is exactly opposite to the three-dimensional scenario \cite{cab, che} where the MF interaction is attractive and the BMF interaction is repulsive. Furthermore, the nonlinear exponent arising from beyond mean-field interaction is quartic in nature \cite{cab}. In our recent work, we have shown that it is possible to reduce the system described in Ref.~\cite{cab} from 3+1 dimension to 1+1 dimension by applying systematic dimension reduction scheme \cite{deb}. It is noteworthy that, BEC in Q1D is experimentally achievable while we know that there is no condensation in 1D system. Thus, a wider investigation of JO in Q1D system poses exciting prospects. 

In mathematical terms, a Q1D system comprises of a NLSE with cubic and quartic nonlinearity with attractive MF interaction (cubic nonlinearity) and repulsive BMF interaction (quartic nonlinearity) while in a 1D system the nonlinear exponents are cubic (MF) and quadratic (BMF). The cubic nonlinearity is positive (repulsive) and the quadratic nonlinearity is negative (attractive). In this article, we plan to study the elementary excitation and Josephson oscillation in such quasi one-dimensional (Q1D) system and analyze the characteristic difference with 1D system \cite{khan}.
However, our main focus will remain in Q1D and we will provide some complementary results of 1D for better quantitative understanding.

Here, we study the JO in a double well potential while the BEC is subjected to mean-field (MF), beyond mean-field (BMF) and three-body (3B) interactions. Recently we have noted that the role of 3B interaction can play crucial role based on the dimensionality \cite{kathapla}. It is worth noting that, about two decades back, quantum liquid was predicted originating from the competition between cubic MF and quintic 3B interaction \cite{bulgac1}. In recent years this competition is also studied from the perspective of flat-top solitons \cite{alotaibi2023unidirectional}. Though these ideas are already existing, here we insert them in a different system domain where we try to capture their role in the JO and ST phases and expect these will provide a more comprehensive picture.

Hence, in this article, we plan to present a systematic analysis of (i) the role of dimensionality by studying Q1D and 1D system; (ii) role of interactions or the competition between MF, BMF and 3B interactions; (iii) role of asymmetry and (iv) imprint of interaction competition and density on Bogoliubov dispersion. The first two objectives are mainly carried out from the perspective of JO in a double well potential. This allows us to extend our study to the third objective where we incorporate asymmetry in the confining potential. In addition we carry out the dispersion analysis via Bogoliubov theory of small perturbation and note the existence of roton-like mode.

The article is arranged in the following way: we elaborate about the physical setup and basic theory in Sec.~\ref{theory}. Then we present our result for symmetric double well potential in Sec.~\ref{result_sym} and extend our analysis for asymmetric potential in Sec.~\ref{assymetry}. Later, we study the Bogoliubov mode in Sec.~\ref{quasi-particle} and we draw our conclusion in Sec.~\ref{conclusion}.

\section{Theoretical Model}\label{theory}
Tunneling of particles through a barrier is one of the fascinating aspects of quantum mechanics which is entirely absent in classical realm of physics. The most common and well studied aspect on tunneling dynamics relates with Josephson effect which suggests tunneling between two macroscopic coherent wave functions. This was first realized when two superconductors were separated by an insulating material \cite{likharev}. The experimental realization of BEC \cite{anderson, davis} had instigated further exploration in the direction of Josephson effect (JE) in atomic BEC \cite{smerzi, gati}. The archetypal way of investigating the JE is to employ a double well potential. One can observe, three distinct regime of tunnelling and correlation can be viewed based on the population imbalance among the wells. At low particle imbalance between the wells, one can observe the plasma oscillation (PO), while at moderate imbalance, Josephson oscillation (JO) and at high imbalance leads to self trapping (ST) \cite{sofia}. The recent realization of quantum liquid \cite{ifb} has further opened up the exciting prospect to observe these dynamical behaviours in light of the BMF interaction \cite{P.wysocki_josephson_2024}.

\begin{figure}[h]
	\hskip -1mm
	\centering
	\includegraphics[width = 1.0\linewidth]{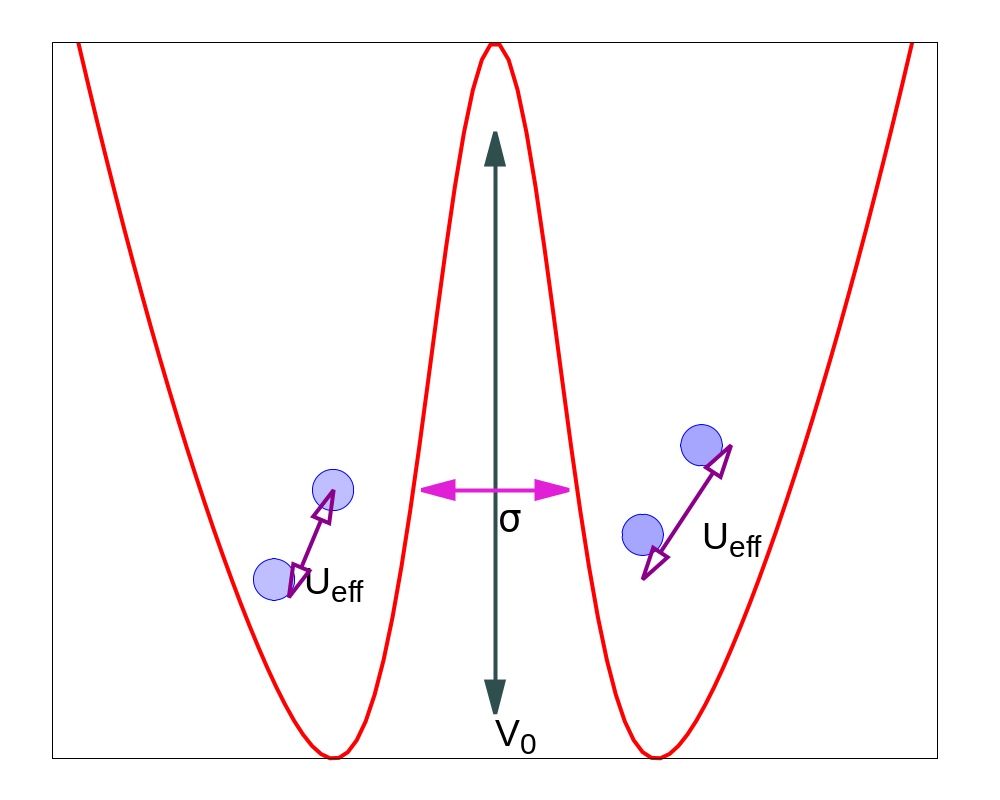}
	\caption{Schematic representation of the physical situation for particles trapped in a double well potential. The blue circles describes the atoms. $U_{eff}$ includes MF, BMF and 3B interactions. $V_0$ is the barrier height.}\label{Double_well_potential}
\end{figure}

Here, we focus on exploring how the effects of the BMF interaction influences the dynamics of small-amplitude oscillations and their frequency at different interaction regimes in Q1D. A shift in these frequencies would serve as a clear and experimentally observable indication of the BMF effect. Hence, we begin our investigation from the Q1D binary Bose mixture in double well potential. We define the double well potential in the following way,
\begin{equation}\label{double_well_potential}
	V(x) = \frac{1}{2} m \omega^{2}_{ho} (x+\Delta x)^{2} + V_{o} e^{- 2 x^{2} / \sigma^{2}},
\end{equation}   
where, $\omega_{ho}$ is the frequency of harmonic trap, $V_{0} $ is the height of potential barrier and $\sigma$ is the width of the barrier and $m$ being the mass of the particle (see Fig.~\ref{Double_well_potential}). $\Delta x$ defines the asymmetry in the double well, however at this moment we set $\Delta x=0$, nevertheless, we will revisit the aspect of asymmetry in Sec.~\ref{assymetry}. 

To study the elementary excitation we can consider the degree of population imbalance in each well. Based on the imbalance and the interaction competition, we expect to observe PO, JO or ST regimes. 

To be precise, we will focus on the Q1D geometry with extended GP equation, which takes the dimensionless form as \cite{debnathjpb, kathapla},
\begin{equation}\label{GPE:BMF}
		i \partial_{t} \psi = \Big[ -\frac{\partial^{2}_{x}}{2} + V(x) + U_{eff}\Big]\psi.  
\end{equation}
Here, $V(x)$ denotes the external confining potential which is double well in our case (as noted in Eq.(\ref{double_well_potential})). $U_{eff}$ defines the effective interaction energy which includes the usual two-body mean field interaction along with BMF and three-body interaction such that $U_{eff}=g_3 |\psi|^{2} + g_4|\psi|^{3} + g_5|\psi|^{4}$. Here, $g_3$ is the strength of the mean-field interaction (cubic nonlinearity) which we will later choose as attractive, $g_4$ describes the strength of the BMF interaction (quartic nonlinearity) and $g_5$ takes into account the three-body interaction strength (quintic nonlinearity). In the later part of our calculation we have chosen BMF and 3B interaction as repulsive. The choice of MF and BMF interactions are inspired from the fact that in Q1D system, the MF interaction is attractive and BMF interaction is repulsive as stated earlier. 
The particle number is related with the normalization condition such that,
\begin{equation}\label{Normalization}
	N = \int |\psi(x)|^{2} dx.
\end{equation} 

Since, the governing GP equation is cast in dimensionless form, $N$ throughout in our calculations should be interpreted as the dimensionless normalization or rescaled particle number of the condensate, rather than reflecting the true physical atom population. Apparently from the figure, we observe minimal change in the wave function, however, it is already been noted that relative small deviation in the wave function may also result in significant dynamical change \cite{P.wysocki_josephson_2024}. Therefore it is more instructive to study the JE which one can also realize experimentally.

\subsection{Two mode approximation}
The familiar prescription to understand the dynamics of two macroscopic wave functions separated by the barrier is via two mode approximation. Here, the wave function is characterized by fractional relative population which is defined as,
\begin{equation}\label{imbalance}
	z(t) = \frac{N_{L}(t) - N_{R}(t)}{N_{L}(t) + N_{R}(t)},
\end{equation}
where, $N_L$ and $N_R$ stands for particle number in the left and right well respectively. It is clear from Eq.(\ref{imbalance}) that when the particle number in each well will be equal then $z(t)=0$ and when all particles are in one well $z(t)=1$. The number of particles in left (right) well is calculated as $ N_{L}(t) = \int_{-\infty}^{0} |\psi(x, t)|^{2} dx$ ($N_{R}(t) = \int_{0}^{\infty} |\psi(x, t)|^{2} dx$). 

Additionally, it is important to study the phase coherence between the particles of the two wells and therefore we define the relative quantum phase difference as,
\begin{equation}\label{Relative_phase}
	\theta(t) = \theta_{R}(t) - \theta_{L}(t)  
\end{equation}
Here, $\theta_L$ ($\theta_R$) defines the phase of the Bosonic collection in the left (right) well. We consider the centre of the well as the reference point to calculate the phase.

We realize that a collection of particles in each well can experience competition between kinetic force, potential energy and different interactions. This can lead to change in particle number in each well. Thus, to understand the quantum dynamics in a double well BEC, imbalance shift (change in $z(t)$) and relative phase shift (change in $\phi(t)$) plays a prominent role. These shifts effects the behaviour of a system, determining whether it goes to Plasma oscillation, Josephson oscillation, or even self trapping. 

Under the two mode approximation, the condensate wave function can be written as,
\begin{equation}\label{Two_mode_approx_ansatz}
	\psi(x, t) = \phi_{L}(x) c_{L}(t) + \phi_{R}(x) c_{R} (t)
\end{equation}  
The functions  $\phi_{L}$ and  $\phi_{R}$ correspond to the spatial modes localized in the left and right wells respectively. The time-dependent complex coefficients $c_{i}(t)$ describe the population and phase dynamics in each mode. The localized modes are computed numerically as $\phi_{L/R} = (\psi_{0} \pm \psi_{1} )/\sqrt{2}$ , where $\psi_{0}$ and $\psi_{1}$ are the lowest symmetric and anti-symmetric stationary solutions of Eq.(\ref{GPE:BMF}). Substituting the two-mode ansatz in Eq.(\ref{Two_mode_approx_ansatz}) into Eq.(\ref{GPE:BMF}) allows us to derive the corresponding dynamical equations.
\begin{widetext}
\begin{eqnarray}
	&&i \partial_{t} \Bigg[ \phi_{L}(x) c_{L}(t) + \phi_{R}(x) c_{R}(t) \Bigg]
	  = \Bigg[ -\frac{\partial_{x}^{2}}{2} + V(x) + g_3|\phi_{L}(x) c_{L}(t) + \phi_{R}(x) c_{R}(t)|^{2}\nonumber\\&& + g_4|\phi_{L}(x) c_{L}(t)+ \phi_{R}(x) c_{R} (t)|^{3} + g_5|\phi_{L}(x) c_{L}(t)+ \phi_{R}(x) c_{R}(t)|^{4} \Bigg](\phi_{L}(x) c_{L}(t) + \phi_{R}(x) c_{R}(t))\nonumber\\\label{left_eq}
\end{eqnarray}
\end{widetext}

While projecting Eq.(\ref{left_eq}) onto $\phi_{L}(x) $, we excluded the cross term (interaction between $\phi_{L}(x) $ and $\phi_{R}(x) $) and focus on the self interaction energy arising from particle in mode $\phi_{L}(x)$ only. Therefore, we project onto the basis state which involves multiplication of each side by $\phi_{L}^{*}$ and integrating over all space which leads to
\begin{widetext}
\begin{eqnarray}
	i \dot{c}_{L}(t) &=& \int dx \phi_{L}^{*}(x) \Bigg[ -\frac{\partial_{x}^{2}}{2} + V(x)\Bigg]\phi_{L}(x) c_{L}(t) + \int dx \phi_{L}^{*}(x) \Bigg[ - \frac{\partial_{x}^{2}}{2} + V(x)\Bigg]\phi_{R}(x) c_{R}(t) \nonumber\\&&+ g_3 \int dx |\phi_{L}(x)|^{4} |c_{L}(t)|^{2} c_{L}(t) + g_4 \int dx |\phi_{L}(x)|^{5} |c_{L}(t)|^{3} c_{L}(t) + g_5\int dx |\phi_{L}(x)|^{6} |c_{L}(t)|^{4} c_{L} (t) \nonumber\\
	i \partial_{t}c_{L} &=& E_{1}c_{L}(t) - J_{0} c_{R}(t) + \bar{g _3} |c_{L}(t)|^{2} c_{L}(t) +  \bar{g_4}  |c_{L}(t)|^{3} c_{L}(t) + \bar{g_5} |c_{L}|^{4} c_{L}(t)\label{Eqn_for_CL}
\end{eqnarray}
\end{widetext}
The equation for $c_{R}(x)$ is analogous to Eq.(\ref{Eqn_for_CL}) which one can easily obtain from Eq.(\ref{Eqn_for_CR}) by projecting Eq.(\ref{left_eq}) onto $\phi_{R}(x) $ and following the above mentioned prescription. The final equation for right well reads as, 
\begin{widetext}
 \begin{eqnarray}\label{Eqn_for_CR}
		i \partial_{t}c_{R} &=& E_{2}c_{R}(t) - J_{0} c_{L}(t) + \bar{g _3} |c_{R}(t)|^{2} c_{R}(t) +  \bar{g_4}  |c_{R}(t)|^{3} c_{R}(t) + \bar{g_5} |c_{R}|^{4} c_{R}(t).
\end{eqnarray}
\end{widetext}

Here, $E_{i} =  \int dx \phi_{i}^{*}(x) \Bigg[ -\frac{\partial_{x}^{2}}{2} + V(x)\Bigg]\phi_{i}(x)$ with $i\in\{1,2\}$ and $J_{0} =-\int dx \phi_{L}^{*}(x) \Bigg[ -\frac{\partial_{x}^{2}}{2} + V(x)\Bigg]\phi_{R}(x)$ are onsite energy of each mode and tunnelling coefficient respectively. $\bar{g _3}$, $\bar{g_4}$ and $\bar{g_5}$ are the nonlinear parameters for MF , BMF and three body interaction which are defined as,

	\begin{eqnarray*}
	    \bar{g _3} &=& g_{3} \int dx |\phi_{L} (x) | ^{4},\\   
		 \bar{g_4} &=& g_{4} \int dx |\phi_{L} (x) |^{5},\\
		\bar{g_5} &=& g_{5} \int dx |\phi_{L} (x) |^{6}
	\end{eqnarray*}
Hence, after numerically calculating $\phi_{L/R}$ from Eq.(\ref{GPE:BMF}) we can obtain $\bar{g_3}$, $\bar{g_{4}}$ and $\bar{g_{5}}$ which later helps us to solve the coupled equation noted in Eqs.(\ref{Eqn_for_CL}, \ref{Eqn_for_CR}).

\subsection{Study of Imbalance}
The schematic in Fig.~\ref{Double_well_potential} describes a situation when particles are trapped in a symmetric double well potential. However, based on the barrier height between the two wells, kinetic and interaction energy of the particles, there is finite probability of the particles to move from one well to another. Experimentally it is possible to allocate equal number of particles in a each well however, the self assembling of the particles due to the interaction competitions can be instructive. It is well understood that over time there is possibility of imbalance in particle numbers inside the wells. Our primary objective in this subsection is to study the excitation dynamics leading from this imbalance.

 
The time dependent complex coefficients $c_j$'s (where $j\in\{L, R\}$) can be described as a combination of amplitude and phase driving functions such that $c_j(t)=\rho_{j} e^{i \theta_{j}}$ while $\rho_j(t)=\sqrt{N_j(t)}$, $N_j(t)$ being the particle number at each well. Now the particles in left and right well can be written in the form of imbalance such that, 
$$ N_{L} = \frac{N}{2} (1 + z), \text{     }  N_{R} = \frac{N}{2} (1 - z) $$ 
where, $N$ is total number of particles or $N = \sum_{j}N_{j}$. 

Substituting the new definition of $c_j(t)$ in Eqs.(\ref{Eqn_for_CL}, \ref{Eqn_for_CR}) along with the above mentioned constrained conditions and applying the definitions from Eqs. (\ref{imbalance}, \ref{Relative_phase}), we obtain a set of coupled equation for $z(t)$ and $\theta(t)$ such that,
\begin{eqnarray}
	\dot{z}&=&-2 J_{0} \sqrt{1 - z^{2}}\sin{\theta}\label{Eq_z_nonscale}\\
	\dot{\theta} &=& \bar{g _3} N z + \frac{2 J_{0} z}{\sqrt{1 - z^{2}}}\cos{\theta} \nonumber\\&& - \frac{1}{2\sqrt{2}}\bar{g_4}N^{3/2} \Big[ (1 - z)^{3/2} - (1 + z)^{3/2} \Big]  \nonumber\\&& - \frac{\bar{g_5}}{4}N^{2} \Big[ (1-z)^{2} - (1 + z)^{2}\Big].\label{Eq_theta_nonscale}
\end{eqnarray}
The variable $z(t)$ and $\theta(t)$ are canonically conjugate variables therefore, $\dot{z}=-\frac{\partial \mathcal{H}}{\partial\theta}$ and $\dot{\theta}=\frac{\partial \mathcal{H}}{\partial z}$ and that leads to the Hamiltonian ($\mathcal{H}$) as, 
 \begin{eqnarray}\label{H_nonscaled}
 	\mathcal{H} &=& \frac{\bar{g _3} N z^{2}}{2}  - 2 \text{} J_{0} \sqrt{1 - z^{2}} \text{ } \cos {\theta} \nonumber\\&& + \frac{1}{5 \sqrt{2}}\bar{g_4} N^{3/2} \Big[ (1 - z)^{5/2} + (1 + z)^{5/2}  \Big] \nonumber\\&& + \frac{\bar{g_5}}{12} N^{2} \Big[ (1-z)^{3} + (1 + z)^{3}   \Big]
 \end{eqnarray}
For convenience of calculation, we re-scale Eq.(\ref{H_nonscaled}) in units of $2J_0$ such that, 
\begin{eqnarray}\label{H_scaled}
	\bar{\mathcal{H}} &=& \frac{\Lambda \text{ } z^{2}}{2} - \sqrt{1 - z^{2}} \text{} \cos{\theta} \nonumber\\&& + \gamma \text{ }  \Big[ (1 - z)^{5/2} + (1 + z)^{5/2}  \Big] \nonumber\\&& + \eta \Big[ (1-z)^{3} + (1 + z)^{3} \Big]
\end{eqnarray}
Here, $\bar{\mathcal{H}} = \mathcal{H} / 2 \text{} J_{0}$,  $\Lambda = \bar{g _3}  N / 2 \text{} J_{0}$, $\gamma = \text{ }\bar{g_4} N^{3/2} / 10 \sqrt{2} J_{0}$ and $\eta$ will be, $\bar{g_5} N^{2} / 24 J_{0}$.

It is important to note here that, if we consider $\bar{g _3} = 0$, then the Hamiltonian boils down to the non-rigid pendulum problem \cite{raghavan1999coherent} while, for $\bar{g _3} \neq 0, \text{and} \bar{g_4} = \bar{g_5} = 0$ , the system described the prominent two mode model \cite{smerzi}. Rewriting Eqs. (\ref{Eq_z_nonscale}) and (\ref{Eq_theta_nonscale}) after scaling by $2J_0$ we obtain, 
 \begin{eqnarray}
 	\dot{z}&=& -\sqrt{1 -z^{2}} \text{} \sin{\theta}\label{Eq_z_scaled}\\
 	\dot{\theta} &=& \Lambda z + \frac{z}{\sqrt{1 -z^{2}}} \cos{\theta} \nonumber\\&& - \frac{5}{2} \gamma \Big[ (1 - z)^{3/2} - (1 + z)^{3/2} \Big] \nonumber\\&& - 3 \eta \Big[ (1-z)^{2} - (1 + z)^{2}\Big].\label{Eq_theta_scaled}
 \end{eqnarray}

As mentioned earlier, depending upon the imbalance, the system can exhibit different phases. This information is assimilated in the knowledge of $\Lambda$ which is proportional to the ratio of the on-site interaction energy and the coupling matrix element. $\Lambda$  acts such a way that one can visualize it like a ``nonlinearity dial''; as you increase it, you shift the system from a linear wave-like state to a nonlinear, localized state. It is invariant under the transformation $\Lambda\rightarrow-\Lambda$ (as $J_{0}\rightarrow -J_{0}$ and $\theta\rightarrow\theta+\pi$). 
In practice $\Lambda$ operates as the control parameter and provide the bridge between JO and ST. 

In the previous investigations, we have noted that $\Lambda$ remained positive as the two body mean-field interaction was repulsive. This has allowed us to obtain a critical value of $\Lambda$ such that $\Lambda_c=\frac{1+\sqrt{1-z_0^2}\cos\theta_0}{z_0^2/2}$ \cite{raghavan1999coherent}. However, in a simple attractive BEC (please see Eq.(\ref{H_scaled}) without BMF and 3B interaction) the first term will always be negative. Now if we choose $\theta_0$ as zero then $\bar{\mathcal{H}}<0$ and one cannot see any ST like phase; nevertheless if $\theta_0=\pi$ then there is a possibility to observe some localization phase as $0<\bar{\mathcal{H}}<1$. However, in presence of BMF and 3B interaction we are actually able to see a ST phase implying $\bar{\mathcal{H}}>1$ for $\pi-$phase mode. Our analysis also reveals that by weakening the MF interaction or sufficiently increasing the BMF interaction \cite{tononi2019quantum}, it is even possible to see ST phase for $\theta_0=0$.


We have already observed that the system step into the self trapping regime whenever $\Lambda$ overcome the value $\Lambda_{c}$ in 1D. It is noted that $\mathcal{H}>-1-2\zeta$ leads to the critical condition leading to self trapping where $\zeta$ denotes the BMF contribution \cite{abdullaev2024beyond}. In the complementary equations to Eqs. (\ref{Eq_z_scaled}, \ref{Eq_theta_scaled}) in 1D if we insert $\theta_{0} = 0$, parameter $\Lambda_{c}$ can be derived, such that $\Lambda_{c} (z_{0}) = 2 (\sqrt{1 - z_0^{2}} + 1) / z_{0}^{2} $ and  for the case where, $\Lambda < \Lambda_{c}$, the system will perform the Josephson oscillation \cite{smerzi}.

\begin{figure}
\includegraphics[width = 1\linewidth]{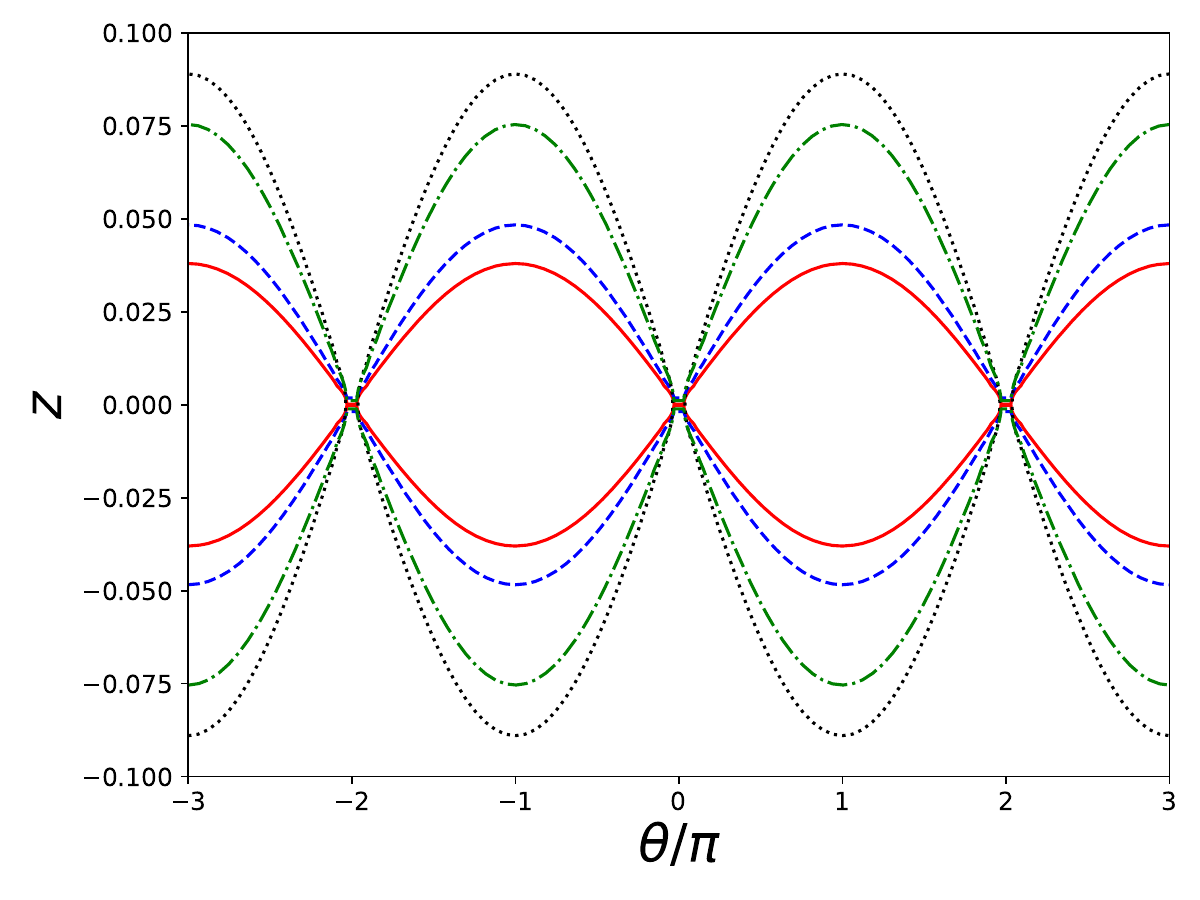}
\caption{ (color online) Figure describes the phase portrait for Q1D system. Solid red curve notes the presence of MF interaction only for initial critical imbalance $z_{0} = 0.038$. Blue dashed curve takes into account MF and BMF for $z_{0} =0.0484$, while black dotted curve depicts all interaction combination i.e. MF, BMF and 3B when we fix the initial imbalance, $z_{0}$ at $0.089$. The green dashed-dotted curve shows the combination of MF and 3B interactions for $z_{0} = 0.0754$.}\label{phase_space}
\end{figure}

We observe in Q1D, the $\pi$ phase mode ($\theta_0=\pi$) captures the transition and the critical condition appropriately. Hence, the critical value of $\Lambda$ from Eq.(\ref{H_scaled}) turns out to be, 
 \begin{eqnarray}\label{lambda_c}
 	\Lambda_{c}(z_{0})  &=& \frac{2}{z_{0}^{2}} \Bigg[ -1 +  \sqrt{1 - z_{0}^{2}} \text{ } \cos {\theta_{0}} \nonumber\\&& + \gamma \Big(  2 -  (1 - z_{0})^{5/2} - (1 + z_{0})^{5/2} \Big)  \Big] \nonumber\\&& + \eta \Big(  2 -(1 - z_{0})^{3} - (1 + z_{0})^{3} \Big)  \Bigg] 
 \end{eqnarray}  
The corresponding phase space diagram is reported in Fig.~\ref{phase_space}. The level crossings in the phase-space are the points of JO to ST transition. The enclosed curves are the regions of JO and the trajectories outside the loop corresponds to the ST regime. We have noted that the critical imbalances are different for different interaction combinations. In presence of MF interaction only, we can keep minimum imbalance at $z_0=0.038$ to see a ST-like phase while in presence of all three interactions we require maximum imbalance as $z_0=0.089$ for JO to ST transition. 

It is also worth noting here that the condition for JO and ST actually reverses in Q1D system. In 1D system (where MF interaction is repulsive) the condition corresponding to JO and ST is $\Lambda<\Lambda_c$ and $\Lambda>\Lambda_c$ respectively. However, in this case a careful calculation reveals that the conditions actually gets reversed which implies ST for $\Lambda_c>\Lambda$ and JO for $\Lambda_c<\Lambda$.
The Josephson oscillation frequency can be extracted from Eqs.(\ref{Eq_z_nonscale}, \ref{Eq_theta_nonscale}), 
\begin{eqnarray}
	{\omega_{J}}_{Q1D} &=& \sqrt{2 J_{0} \Big(  2 J_{0} + \bar{g _3} N + \frac{3}{2 \sqrt{2}} \bar{g_4} N^{3/2} + \bar{g_5} N^{2} \Big)}.\nonumber\\\label{Eq_omegaJ_Q1D}
\end{eqnarray}

	\begin{figure*}[ht]
		\centering
		\includegraphics[width = 0.98\linewidth]{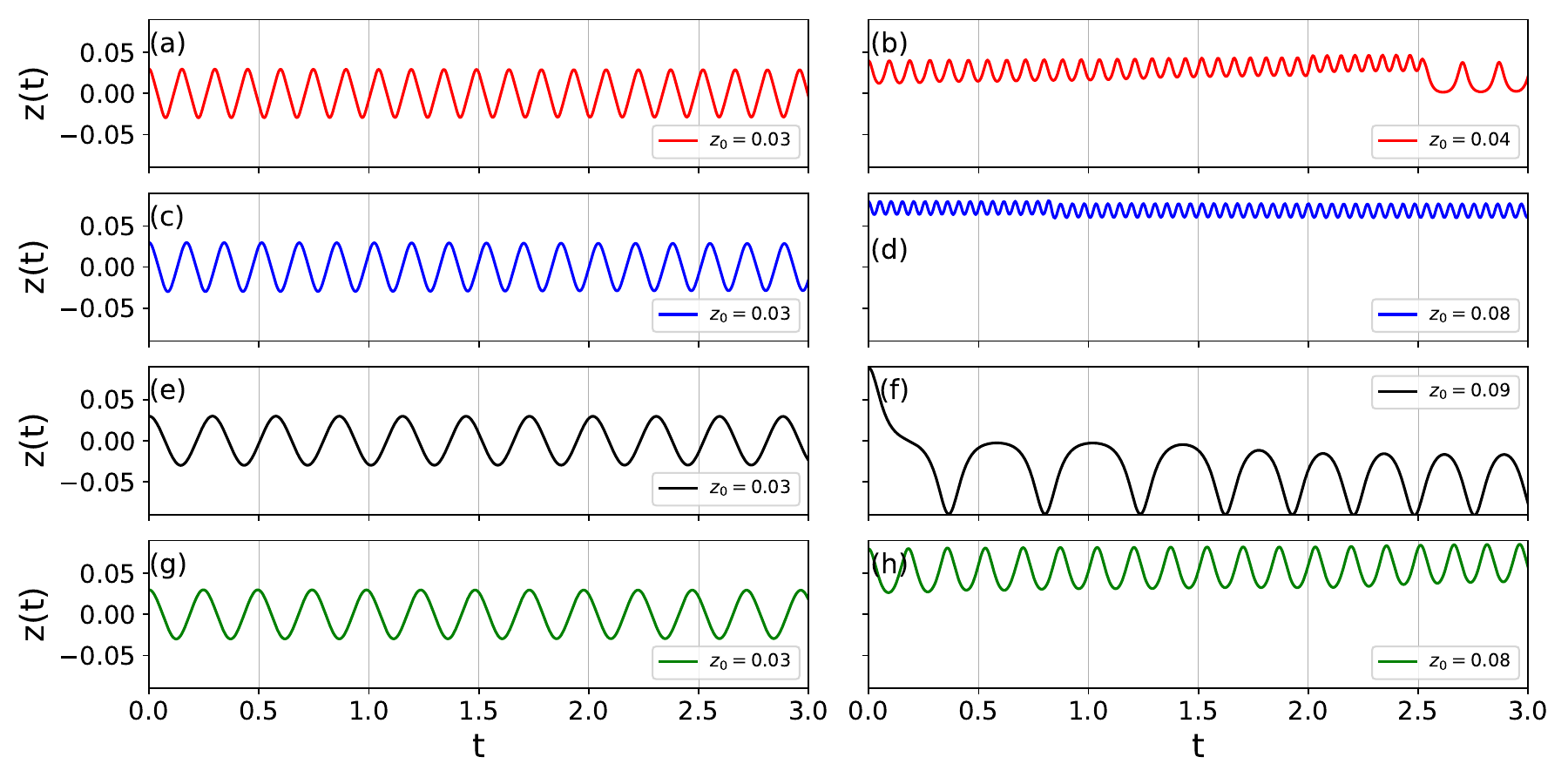}
		\caption{(color online) Description of various combination of interaction strength in Q1D system, which describe how the system reacts from Josephson to self trapping regime by accounting the value of $\Delta x = 0$. In (a) and (b) only MF is considered while in  (c), (d) combination of MF and BMF correction is taken into account. (e) and (f) is prepared when all the interaction term MF + BMF + 3B are present and (g), 
(h) describes only MF and 3B interaction. Here, $g_3=-0.6$, $g_4=0.01$ and $g_5=0.009$. All these values are in arbitrary units. $N$ is taken as $11.5$.}\label{Plot_Q1D}
	\end{figure*}

Till now, we have talked about the mathematical implementation of imbalance and relative phase in higher order correction in Q1D system. By which we are able to settle down the expression for Josephson frequency described in Eq.(\ref{Eq_omegaJ_Q1D}). In similar manner we can get the expression for 1D system \cite{P.wysocki_josephson_2024, abdullaev2024beyond}. In 1D system, the equation of motion can be defined as, 
\begin{eqnarray}
	i \partial_{t} \psi = \Big[ -\frac{\partial^{2}_{x}}{2} + V(x) + U_{eff}\Big]\psi,\label{GP1D}
\end{eqnarray}
where $U_{eff}=g_3 |\psi|^{2} + g_2|\psi| + g_5|\psi|^{4}$. Here, $g_2$ describes the BMF interaction strength while 3B interaction is taken care off by the quintic nonlinearity with interaction strength $g_5$. $g_3$ is the usual MF term. We can again reconstruct the coupled population imbalance and phase imbalance equation, analogous to Eqs.(\ref{Eq_z_scaled}, \ref{Eq_theta_scaled}), such that,
 \begin{eqnarray}
	\dot{z}&=&-\sqrt{1 -z^{2}} \text{} \sin{\theta}\label{Eq_z_scaled_1D}\\
	\dot{\theta} &=& \Lambda z + \frac{z}{\sqrt{1 -z^{2}}} \cos{\theta} \nonumber\\&& + \frac{3}{2} \zeta \Big[ (1 - z)^{1/2} - (1 + z)^{1/2} \Big] \nonumber\\&& - 3 \eta \Big[ (1-z)^{2} - (1 + z)^{2}\Big]\label{Eq_theta_scaled_1D}
\end{eqnarray}
Here, $\Lambda = \bar{g _3}  N /2J_{0}$, $\zeta = \bar{g _2} \sqrt{2 N}/6 J_{0}$ and $\eta = \bar{g_5} N^{2} / 24 J_{0}$. Therefore, the condition for self trapping ($\Lambda > \Lambda_{c}$) regime and Josephson oscillation ($\Lambda < \Lambda_{c}$) frequency can be written as,
 \begin{eqnarray}
	\Lambda_{c}(z_{0})  &=& \frac{2}{z_{0}^{2}} \Bigg[ 1 -  \sqrt{1 - z_{0}^{2}} \text{ } \cos {\theta_{0}} \nonumber\\&& - \zeta \Big(  2 -  (1 - z_{0})^{3/2} - (1 + z_{0})^{3/2} \Big)  \Big] \nonumber\\&& + \eta \Big(  2 -(1 - z_{0})^{3} - (1 + z_{0})^{3} \Big)  \Bigg] \\
	{\omega_{J}}_{1D} &=& \sqrt{2 J_{0} \Big(  2 J_{0} + \bar{g _3} N +  \bar{g_2} \sqrt{N/2} + \bar{g_5} N^{2} \Big)}\nonumber\\\label{Eq_omegaJ_1D}
\end{eqnarray} 
The nonlinear parameters, $\bar{g_3}$ , $\bar{g _2}$ and $\bar{g_5}$, in 1D is defined as,
\begin{eqnarray*}
	\bar{g_3}&=&g_{3} \int dx |\phi_{L} (x) |^{4},\\
	\bar{g_2}&=&g_{2} \int dx |\phi_{L} (x) |^{3},\\
	\bar{g_5}&=&g_{5} \int dx |\phi_{L} (x) | ^{6}
\end{eqnarray*}
These parameters can be calculated from the dynamical equation described in Eq.(\ref{GP1D}).

\section{Result Analysis for Symmetric Double Well Potential}\label{result_sym}
In this section we report our results when the bosons are in a symmetric double well (SDW) potential in Q1D and 1D.

For our numerical calculation we use fixed values for $g_{3}$, $g_{2}$, $g_{4}$ and $g_{5}$ in both the systems. However, we choose the arbitrary values such a way that $|g_3|>|g_2|,\,|g_4|,\,|g_5|$. In all of our calculation we have used $g_5>0$ however, we have observed that the change in nature of 3B interaction (i.e., from repulsive to attractive) does not make any significant change in our main outcomes as 3B interaction strengths are chosen to be relatively small compared to the MF and BMF interactions. In precise, the interactions strengths used in Q1D (1D) are as follows, $g_{3} = - 0.6 (0.6)$, $g_{4} (g_{2}) = 0.01 (-0.1)  $ , $g_{5} = 0.009$ respectively. 
These values are in arbitrary units nevertheless our objective is to have a set of parameter values which can be used for computational purpose across the systems consistently. 

We have also considered the MF limit for our study by switching off the BMF and 3B interaction terms ($g_{4} = g_{5} = 0 $), which suggests legitimate comparison (particularly for 1D system) with conventional MF description of a Bose-Einstein condensate Josephson junction. Our results are consistent when we use $\Lambda\sim15$ as suggested by the Ref.\cite{gati}. For $Rb^{87}$, $\Lambda\sim15$ is determined for the experimental setup  of trap frequencies of $\omega_{x} = 2 \pi \times 78$ Hz, $\omega_{y} = 2 \pi \times 66$ and $\omega_{z} = 2 \pi \times 90$, with gravity acting of y-direction with actual particle number $1150\pm150$ with barrier height of $2 \pi \times 63 $ Hz. The barrier height is responsible for the splitting of the condensate. The $s-$wave scattering length ($a_s$) is considered as $5.31\,nm$ which is equivalent to $\sim100a_B$, where $a_B$ is the Bohr radius. In our calculation $g_3$ is actually $2a_s/a_B$ and $N=11.5$. Appropriate mapping with the experimental data suggests $g_3\sim2$ while we have used $g_3=0.6$. Hence, our choice of $g_3$ is about one-third of Ref.~\cite{gati}. However, it must be noted that, the current analysis covers a much wide spectrum of interaction parameters along with different dimensionality and asymmetry. The current choice of parameters appeared most suitable to study these different aspects. Being consistent with these numbers, our calculation reveals both dynamical regime, namely, JO and nonlinear ST, which are also susceptible to the initial imbalance ($z_0$). 

Since, BMF interaction strength is expected to be weak thus we have considered it one order of magnitude lower as compared to MF interaction strength.
In our analysis, the total number of particles have been normalized to $N$. We have presented plots for $N=11.5$, however, later we have also examined the role of the norm. Within this framework, we have observed that both the BMF correction and the 3B interaction term make considerable contribution to the dynamics of Josephson oscillations. 

In Fig.~\ref{Plot_Q1D}, the left panel describes the Josephson regime while the right panel depicts the self trapping regime ($\langle z\rangle\neq 0$) \cite{sofia} for different interaction competitions in Q1D system. Furthermore, extreme upper plots (in red) is accounting only with MF interactions, below that (in blue) for MF and BMF, second lower (in black) having all the interaction i.e MF, BMF and 3B. In last (in green) having MF and 3B coupling strengths.
The numerical calculation of $J_0$ and $\bar{g_3}$ allows us to calculate $\Lambda$. It is important to highlight, that while the BMF and 3B terms strongly affect the population dynamics and oscillation patterns, the tunnelling coefficient itself does not change across these scenarios. The tunnelling coefficient is determined exclusively by the structure of the external double-well potential and remains fixed as long as the potential barrier and trap geometries are unchanged. We can also calculate the critical $\Lambda$ or $\Lambda_c$ from Eq.(\ref{lambda_c}) for trial values of $z_0$. A systematic calculation and the phase-space analysis (see Fig.~\ref{phase_space}) suggests the precise values of initial imbalance ($z_{0}$) which can lead us from JO to ST regime. 

From Eq.(\ref{H_scaled}) we numerically evaluate $\Lambda$ as $-2762.87$ for $N = 11.5$. We now calculate $\Lambda_c$ using Eq.(\ref{lambda_c}) with trial values of $z_0$. If $\Lambda_c$ value is less than $\Lambda$ we expect to observe JO while if it is more than $\Lambda$, ST will emerge. Since $\Lambda_c$ is a function of $z_0$ hence, we fix some $z_0$ as initial condition to solve the coupled Eqs.(\ref{Eq_z_scaled}, \ref{Eq_theta_scaled}) numerically. For appropriate choice of $z_0$ we can traverse from JO to ST regime and we report this in Fig.~\ref{Plot_Q1D}. 

In Fig.~\ref{Plot_Q1D} (a) and (b) we note Josephson oscillation for initial imbalance $z_{0} = 0.03$ (while $\Lambda_{c} = -4443.4$) and ST is noted at $z_{0} = 0.04 (\Lambda_{c} = -2499)$. It can be seen clearly that these values of initial imbalance obeys the the condition of JO ($\Lambda > \Lambda_{c}$) and ST ($\Lambda < \Lambda_{c}$) perfectly. Similarly, numerically obtained $\Lambda$ value of the system when it is subjected to MF and BMF interaction is evaluated as $-1843.11$. When we take into account MF, BMF and 3B interaction together then $\Lambda=-603$ while in presence of only MF and 3B interaction $\Lambda=-783.14$. In the figure, we have noted that, JO condition for (c), (e) and (g) with $z_{0} = 0.03$ ($\Lambda_{c}= -2637$), $z_{0} = 0.03$ ($\Lambda_{c} =  -725.2$ ) and $z_{0} = 0.03$ ($\Lambda_{c} = -897.3$) respectively are satisfying the above mentioned conditions as well. Similarly, ST condition for (d), (f) and (h) are $z_{0} = 0.05$ ($\Lambda_{c} = -1737$), $z_{0} = 0.09$ ($\Lambda_{c} = -594.0$) and $z_{0} = 0.08$ ($\Lambda_{c} = -706.0$) respectively. For JO using $z_0=0.03$, our $\Lambda_c$ is always less than $\Lambda$ thus we report the JO for this fixed initial imbalance.
	
\begin{figure}[h]
	\centering
	\includegraphics[width = 1.\linewidth]{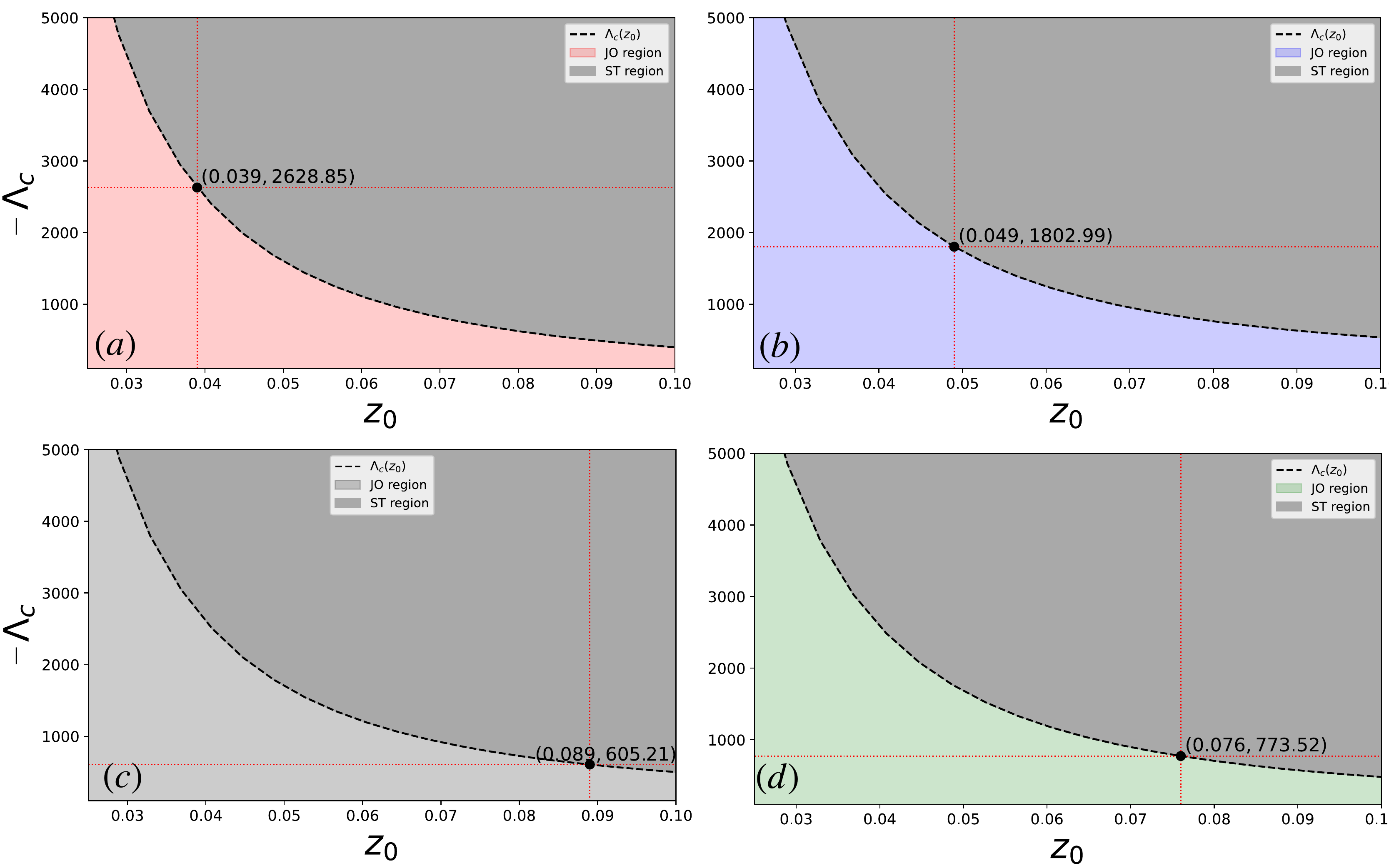}
	\caption{(Color online) The figure shows the phase plot of $\Lambda_{c}$ vs $z_{0}$ for all possible combinations of interaction strengths. (a) Represents the contribution of the mean-field (MF) interaction only. (b) Shows the combined effect of MF and beyond-mean-field (BMF) interactions. (c) Depicts the case including MF, BMF, and 3B interactions. (d) Corresponds to the combination of MF and 3B interactions.}\label{phase_separation}
\end{figure}

In Fig.~\ref{phase_separation} we have summarized the variation of $\Lambda_c$ with $z_0$ using Eq.\ref{lambda_c}. Fig.~\ref{phase_separation}(a) describes the effect only when MF interaction is present while (b) takes into account both MF and BMF interaction. In (c) all three interactions are present and in (d) MF and 3B interaction competes. The figure clearly demarcates the JO and ST boundary and further shows the progressive increase of $z_0$ when we have more interaction competition. It also suggests that, even though 3B effect may appear minuscule but in determining the phase boundary it has more pronounced effect than the BMF interaction. 

It is also fascinating to see in Fig.~\ref{phase_separation}, that on separating the JO and ST regime the global phase boundary almost unaltered under the different interaction regimes. The transition that we have seen from JO to ST is decided by the separatrix in phase space, which is elaborated in ref.\cite{smerzi} and phase boundary is determined by the necessary balance between tunnelling and order of interactions. Although, higher order interactions (proportional to $N^{3/2}$ and $N^{2}$) are small, still they contribute to the total interaction energy which becomes more evident near the transition point. The increased sensitivity of the nonlinear system responds noticeably to the small perturbation resulting sufficient shift in the dynamical behaviour of the system  \cite{strogatz2001nonlinear}. This explain the shift the values of $z_{{0}_c}$ and $\Lambda_{c}$ due to the inclusion of high order correction like BMF and 3B albeit the global boundary of phase transition remain almost unchanged. Similar kind of behavior has also been reported for quantum droplet system \cite{petrov1}, where weak BMF interaction affect the stabilization of the system.
In a nutshell, it can be commented that the interaction competition re-calibrates the fluid's total interaction energy without altering the underlying stability criteria.

\begin{figure*}
	\centering
	\includegraphics[width = 0.95\linewidth]{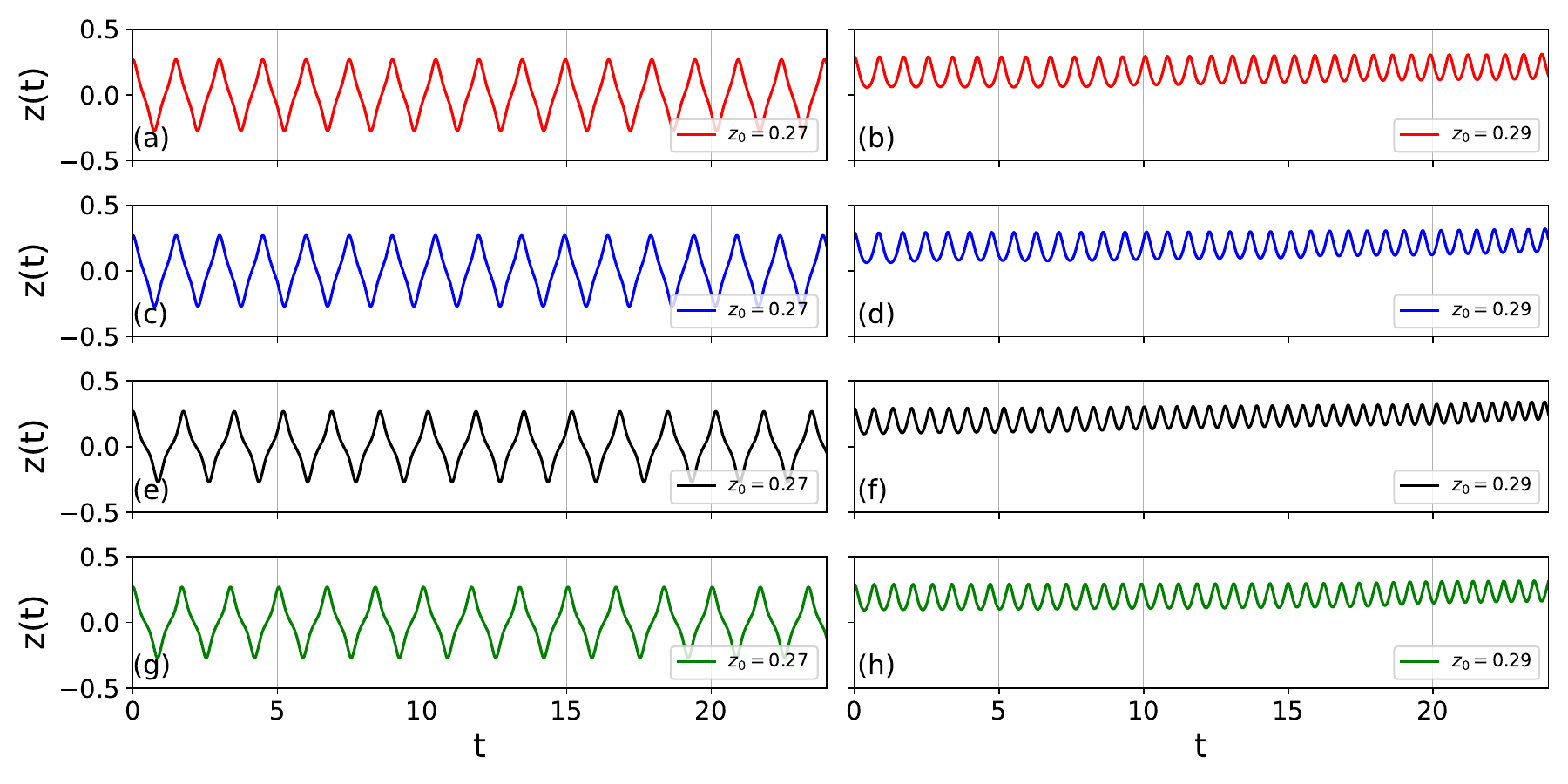}
	\caption{(color online) Description of various combination of interaction strength in 1D system, which describe how the system reacts from Josephson to self trapping regime by taking the value of $\Delta x = 0$. In (a) and (b) only MF is considered while in  (c), (d) combination of MF and BMF correction is taken into account. (e) and (f) is prepared when all the interaction term like MF, BMF and 3B are present and (g), 
(h) describes only MF and 3B interaction. Here, $g_3=0.6$, $g_2=-0.01$ and $g_5=0.009$. All these values are in arbitrary units. $N$ is taken as $11.5$.}\label{Plot_1D}
\end{figure*}

For 1D system we have followed the same procedure and noted down the critical values of $\Lambda$ for $N=11.5$. At first, we focus only on MF interaction. We have observed JO for initial imbalance of $0.27$ while ST for $z_0=0.29$. This is noted in Fig~\ref{Plot_1D} (a) and (b). The choice of $z_0$ is heavily dependent on $\Lambda$ calculation which we obtain as $48.40$. Therefore it is found out that $z_{0} = 0.27$ leads to $\Lambda_{c} = 50.0$ suggesting favourable condition for JO ($\Lambda < \Lambda_{c}$) along with ST ($\Lambda > \Lambda_{c}$) where $z_{0} = 0.29$ leads to $\Lambda_{c} = 46.5)$. Similarly, for all other combinations we have defined $z_{0}$ values to observed JO and ST regime. We fixed $z_0=0.27$ and $0.29$ to report our results for all interaction permutations. 
When the system is subjected to MF and BMF interaction, $\Lambda  = 48.46$ while for the above mentioned population imbalance, $\Lambda_{c}$ will be $49.8$ and $46.3$ respectively for JO and ST regime (see Fig.~\ref{Plot_1D} (c) and (d)). Consequently when we take into account MF, BMF and 3B interaction, $\Lambda = 47.49$ and the critical $\Lambda$ for JO and ST turns out as $49.3$ and $45.4$ respectively (see Fig.~\ref{Plot_1D} (e) and (f)). In the last combination where only MF and 3B interactions are present, $\Lambda$ is $47.53$. The critical $\Lambda$ used to capture the JO and ST can be worked out as $\Lambda_{c}=$ $49.5$ and $45.7$ respectively (Fig.~\ref{Plot_1D} (g) and (h)).

\begin{figure}
\centering
		\includegraphics[width = 1.2\linewidth]{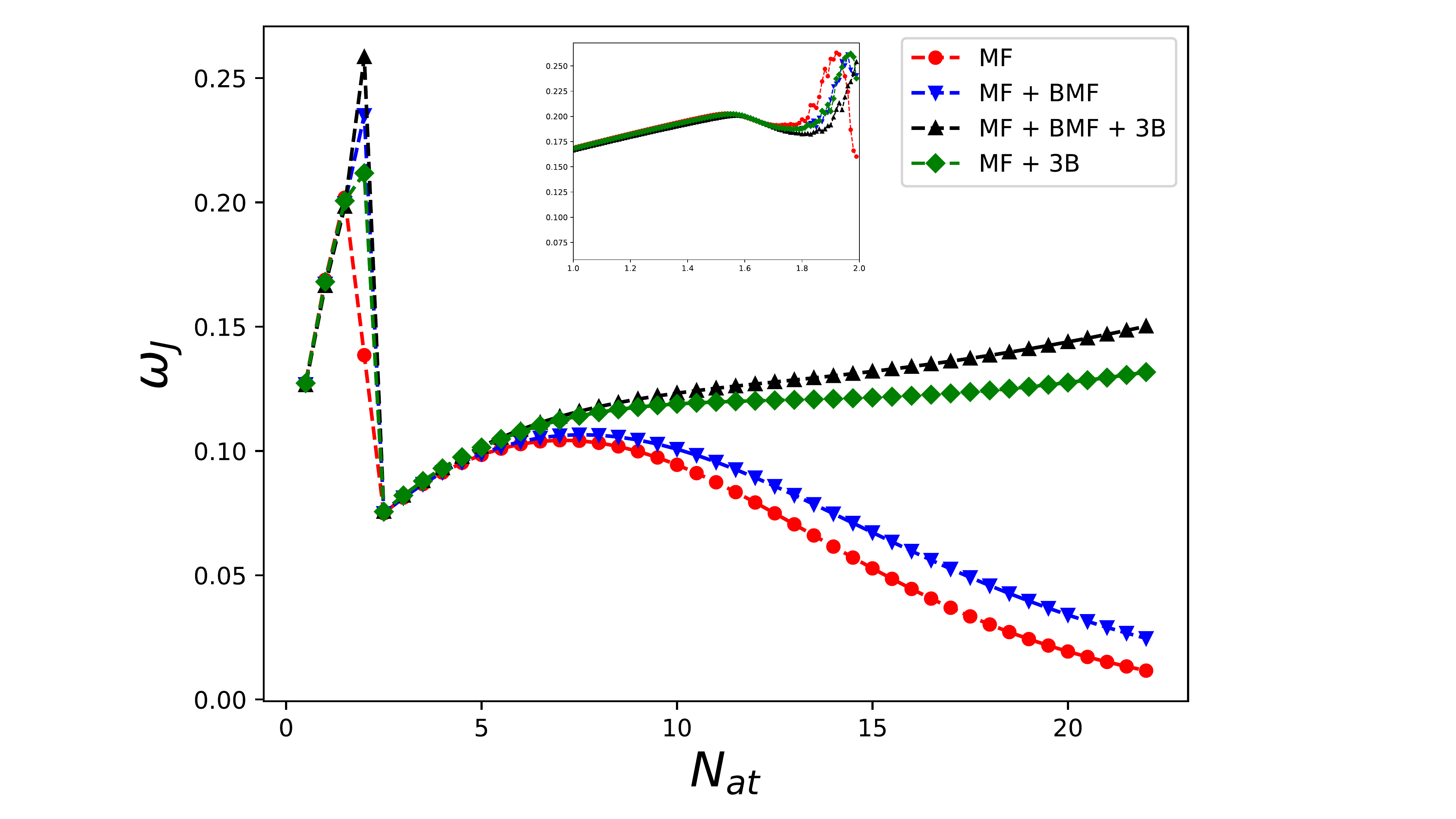}
		\caption{(color online) Variation of Josephson frequency in Q1D system at a fixed imbalance $z_{0} = 0.03$ with variation of particle number. It can be clearly noted that the addition of BMF and 3B interaction actually flattens the non-monotonic nature of the frequency. For calculation purpose, we have used $g_3=-0.6$, $g_4=0.01$ and $g_5=0.009$. All the parameters are in arbitrary units. In the inset we plot the same quantities when normalization is restircted between $1$ to $2$.}\label{fig:WJ_Q1D}
\end{figure}

It is also fascinating to explore the role of normalization on Josephson frequency in Q1D system. Our findings are reflected in Fig.~\ref{fig:WJ_Q1D}. We have prepared the plot from Eq.(\ref{Eq_omegaJ_Q1D}) after numerically evaluating the interaction components from Eq.(\ref{GPE:BMF}). In the figure the blue dashed-solid circle depicts the Josephson frequency when subjected only to MF interaction. Effect of BMF interaction is reflected in red dashed-inverted solid triangle curve while green dashed-solid triangle curve includes all interaction components. The black dashed-solid square takes into account MF and 3B interaction only. We observe relatively large fluctuation of frequency at low normalization. Therefore, we add an inset in the figure where we have depicted the frequency variation at relatively small normalization ($N=1$ to $2$) with smaller grid size calculation. 
The inset reveals that at very low $N$ there is not much frequency variation for different interaction competition, however at around $N\sim 2$ the fluctuation becomes noticeable. Afterwards the frequencies again settles to a minima at $N\sim 2.5$. Later, we observe a non-monotonic behavior  of Josephson frequency in presence of MF as well as MF and BMF interactions. One must note that similar behavior was also noted in Ref.~\cite{P.wysocki_josephson_2024} for 1D system. Also, we have noted that the Josephson frequency follows similar trend as reported in Ref. \cite{abdullaev2024beyond} when we study it as a function of MF interaction. However, in Fig.~\ref{fig:WJ_Q1D}, we additionally observe that the inclusion of 3B interaction linearizes the frequency. It is clear from the figure that at low particle number there is hardly any effect of 3B interaction while, 3B effect becomes more dominant when we increase the particle number as presence of more particle ehnaces the 3B effect. 

\section{Asymmetric Double Well Potential}\label{assymetry}
Since in SDW, the JO frequency variation is relatively small we extend our study to asymmetric double well (ADW) potential hoping that an asymmetry or a titled lattice (from experimental perspective) may yield more pronounced effect of the interaction competition. As our primary focus is Q1D system, we present the detailed calculation and result for Q1D and add only the important formulae for 1D, leaving out further details for brevity. A detailed discussion of asymmetric double well potential for a 1D system can be obtained in Ref.~\cite{raghavan1999coherent} (albeit without the BMF interaction).

\begin{figure}
\includegraphics[scale=0.25]{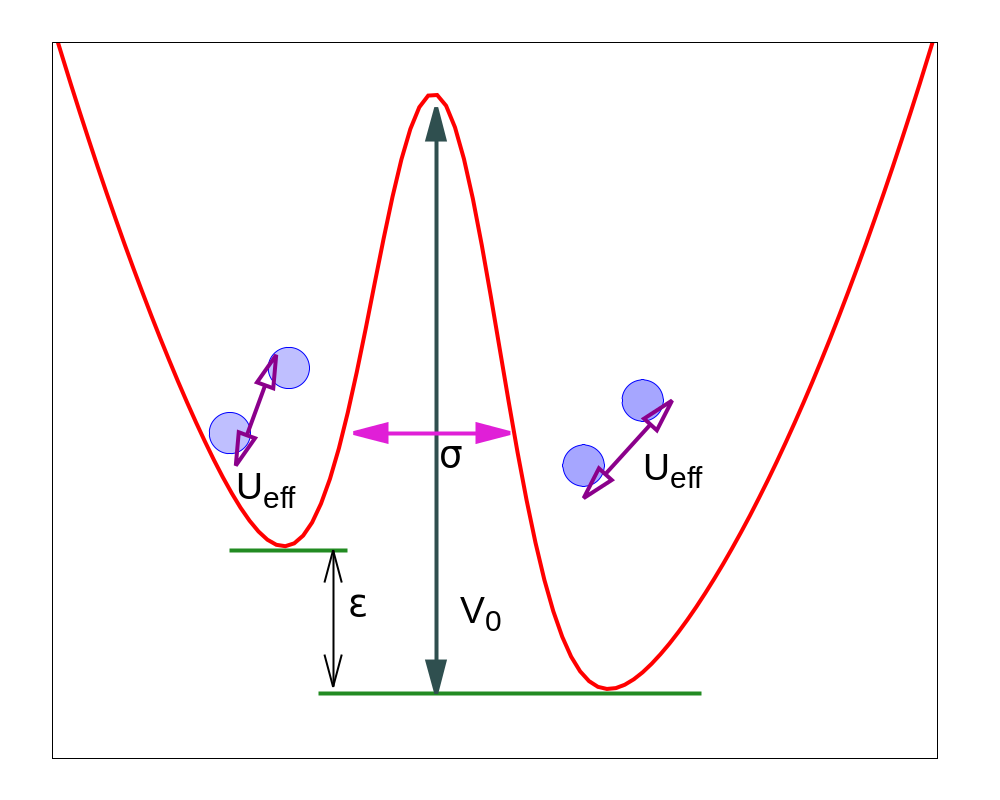}
\caption{Schematic representation of the physical situation for particles trapped in an asymmetric double well potential. The blue circles describes the atoms. $U_{eff}$ includes MF, BMF and 3B interactions. Here, $\Delta x=1$.}\label{ADW}
\end{figure}

Hence, in this section we introduce a level difference (as described in Fig.~\ref{ADW}) in the potential landscape and study its effect. In this direction the first requirement now is to modify the equation of population and phase imbalance to incorporate the asymmetry. Later we summarize our JO result for Q1D. 
\begin{figure*}
	\centering
	\includegraphics[width = 0.98\linewidth]{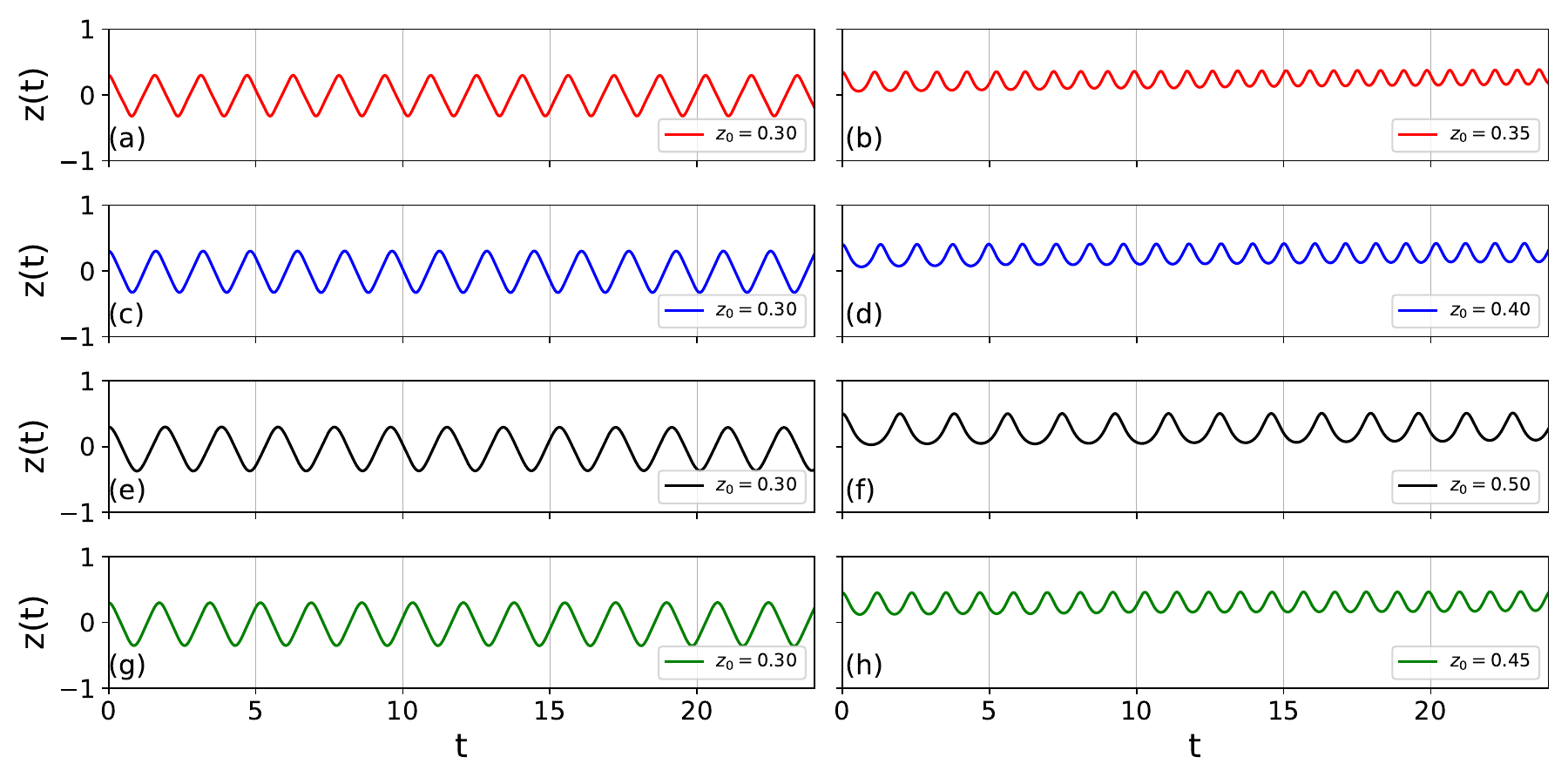}
	\caption{(color online) Description of various combination of interaction strength in Q1D system, which describe how the system reacts from Josephson to self trapping regime. Therefore, the asymmetry is introduced by setting the value of $\Delta x = 1$. In (a) and (b) only MF is considered while in  (c), (d) combination of MF and BMF correction is taken into account. (e) and (f) is prepared when all the interaction term and (g), 
(h) describes only MF and 3B interaction. The interaction parameter values are same noted in  Fig.~\ref{Plot_Q1D}}\label{Plot_Q1D_asym}
\end{figure*}

To incorporate the asymmetry we consider  $\Delta x = 1.0 $ in the potential (see Eq. (\ref{double_well_potential} and Fig.~\ref{ADW}). In this situation it is obvious that by applying the two mode method, the corresponding energies will not be equal (i.e., $E_{1} \neq E_{2}$).  However, the modified two mode equation then reads (after appropriate scaling by $2J_0$),
 \begin{eqnarray}
\dot{z}&=&-\sqrt{1 -z^{2}} \text{} \sin{\theta}\label{Eq_z_scaled_asym}\\
\dot{\theta} &=& \bar{E} + \Lambda z + \frac{z}{\sqrt{1 -z^{2}}} \cos{\theta} \nonumber\\&& - \frac{5}{2} \gamma \Big[ (1 - z)^{3/2} - (1 + z)^{3/2} \Big] \nonumber\\&& - 3 \eta \Big[ (1-z)^{2} - (1 + z)^{2}\Big]\label{Eq_theta_scaled_asym}
\end{eqnarray}
Here, $\bar{E} = \Delta E / 2 J_{0}$ and $\Delta E=E_1-E_2$, rest of the parameters are same as symmetric case. The condition for self trapping now gets modified and it reads,
 \begin{eqnarray}
	\Lambda_{c}(z_{0})  &=& \frac{2}{z_{0}^{2}} \Bigg[- \bar{E} z -  1 + \sqrt{1 - z_{0}^{2}} \text{ } \cos (\theta_{0}) \nonumber\\&& +\gamma \Big(  2 -  (1 - z_{0})^{5/2} - (1 + z_{0})^{5/2} \Big)  \Big] \nonumber\\&& + \eta \Big(  2 -(1 - z_{0})^{3} - (1 + z_{0})^{3} \Big)  \Bigg], 
\end{eqnarray}
and the Josephson frequency can now be noted as,
\begin{eqnarray}\label{Eq_omegaJ_Q1D_asym}
	{\omega_{J}}_{Q1D} &=& \sqrt{2 J_{0} \Big(  2 J_{0} + \bar{g _3} N + \frac{3}{2 \sqrt{2}} \bar{g_4} N^{3/2} + \bar{g_5} N^{2} + \Delta E / z \Big)}\nonumber\\
\end{eqnarray} 


As earlier, we extend the same calculation for 1D system and the modified imbalance equations are noted as,
\begin{eqnarray}
	\dot{z}&=&-\sqrt{1 -z^{2}} \text{} \sin{\theta}\label{Eq_z_scaled_asym_1D}\\
	\dot{\theta} &=& \bar{E} + \Lambda z + \frac{z}{\sqrt{1 -z^{2}}} \cos{\theta} \nonumber\\&& + \frac{3}{2} \zeta \Big[ (1 - z)^{1/2} - (1 + z)^{1/2} \Big] \nonumber\\&& - 3 \eta \Big[ (1-z)^{2} - (1 + z)^{2}\Big],\label{Eq_theta_scaled_asym_1D}
\end{eqnarray}
correspondingly, the equation for critical imbalance and Josephson frequency reads,
\begin{eqnarray}
		\Lambda_{c}(z_{0})_{1D}  &=& \frac{2}{z_{0}^{2}} \Bigg[-\bar{E} z + 1 -  \sqrt{1 - z_{0}^{2}} \text{ } \cos{\theta_{0}} \nonumber\\&& - \zeta \Big(  2 -  (1 - z_{0})^{3/2} - (1 + z_{0})^{3/2} \Big)  \Big] \nonumber\\&& + \eta \Big(  2 -(1 - z_{0})^{3} - (1 + z_{0})^{3} \Big)  \Bigg] \\
		{\omega_{J}}_{1D} &=& \sqrt{2 J_{0} \Big(  2 J_{0} + \bar{g _3} N +  \bar{g_2} \sqrt{N/2} + \bar{g_5} N^{2}  + \Delta E / z \Big)}\nonumber\\\label{Eq_omegaJ_asym_1D}
\end{eqnarray}

Now we are in a position to comment on the role of asymmetry in double well and its effect on the transition between JO to ST. We report the JO and ST regimes in Fig.~\ref{Plot_Q1D_asym}. 
In the figure we follow the same prescription as previous, so that $\Lambda$ comes out as $-30.972$ when we take into account only MF interaction. At $z_{0}=0.3$ we find $\Lambda_{c} =-41.3$ which satisfies the condition for JO. At $z_{0} = 0.35$, $\Lambda_{c}$ turn out as $-29.8$ implying the ST regime can be viewed. We have reported this in Fig.~\ref{Plot_Q1D_asym}(a) and (b). When we take into account both MF and BMF interactions, $\Lambda  = -27.1955$ so that $-45.3$ and $-26.4$ are the value of $\Lambda_{c}$ for JO and ST corresponding to $z_0=0.3$ and $0.4$ respectively (Fig.~\ref{Plot_Q1D_asym} (c) and (d)). When all interaction starts playing a role (implying presence of MF, BMF and 3B) the value of $\Lambda$ is calculated as $-19.1675$ consequently $\Lambda_{c}=-46.1$ and $-19.0$ corresponding to $z_0=0.3$ and $z_0=0.5$ leading to the observation of JO and ST regimes which is displayed in Fig.~\ref{Plot_Q1D_asym} (e) and (f). In presence of only MF and 3B interaction $\Lambda$ is noted as $-22.68$ while $\Lambda_{c}$ for JO and ST are $-44.4$ ($z_0=0.3$) , $-20.8$ ($z_0=0.45$) respectively (see Fig.~\ref{Plot_Q1D_asym} (g) and (h)).

\begin{table}
	\centering
	\caption{Calculated Josephson frequency for symmetric and asymmetric case based on the observations of Fig.~\ref{Plot_Q1D}, \ref{Plot_1D} and \ref{Plot_Q1D_asym}}\label{Table:symmetric}
\vskip 0.2cm
	\begin{tabular}{c@{\hskip 5pt} c@{\hskip 5pt} c@{\hskip 5pt} c@{\hskip 5pt} c@{\hskip 5pt} c }
		\hline \hline
		Interaction  & $(\omega_{Q1D})_{SDW}$ & $(\omega_{1D})_{SDW}$ & $ (\omega_{Q1D})_{ADW}  $ \\  
	     \\ \hline \\
		MF & 0.08344 & 0.47132 & 1.5376  \\ [0.3cm]
		
		MF + BMF  & 0.09378 & 0.47044 & 1.46671  \\[0.3cm]
		
		MF + BMF + 3B & 0.12778 & 0.48981 & 1.2782\\ [0.3cm]
		
		MF + 3B & 0.11996& 0.49062& 1.3200 \\
		\hline \hline
	\end{tabular}
	\end{table}

We have summarized our results in Table.\ref{Table:symmetric} where we have written down the Josephson frequency for symmetric Q1D, 1D and asymmetric Q1D. The Josephson frequencies were noted corresponding to $z_0=0.03$, $0.27$ and $0.3$ for Q1D, 1D and asymmetric Q1D system respectively. We can clearly note the frequency variation as a result of different interaction competition. Since PO regime appears at very low imbalance, thus in symmetric Q1D case, PO can only be viewed at a extremely low imbalance. The JO to ST transition is observed in the vicinity at $z_0=0.038$. However, in 1D and asymmetric Q1D, at moderate critical imbalance we can notice JO to ST transition. This allows us to believe it is relatively easier to observe both PO, JO and ST phases in 1D and asymmetric Q1D systems. 

\begin{figure}[h]
		\centering
		\includegraphics[width=0.85\linewidth]{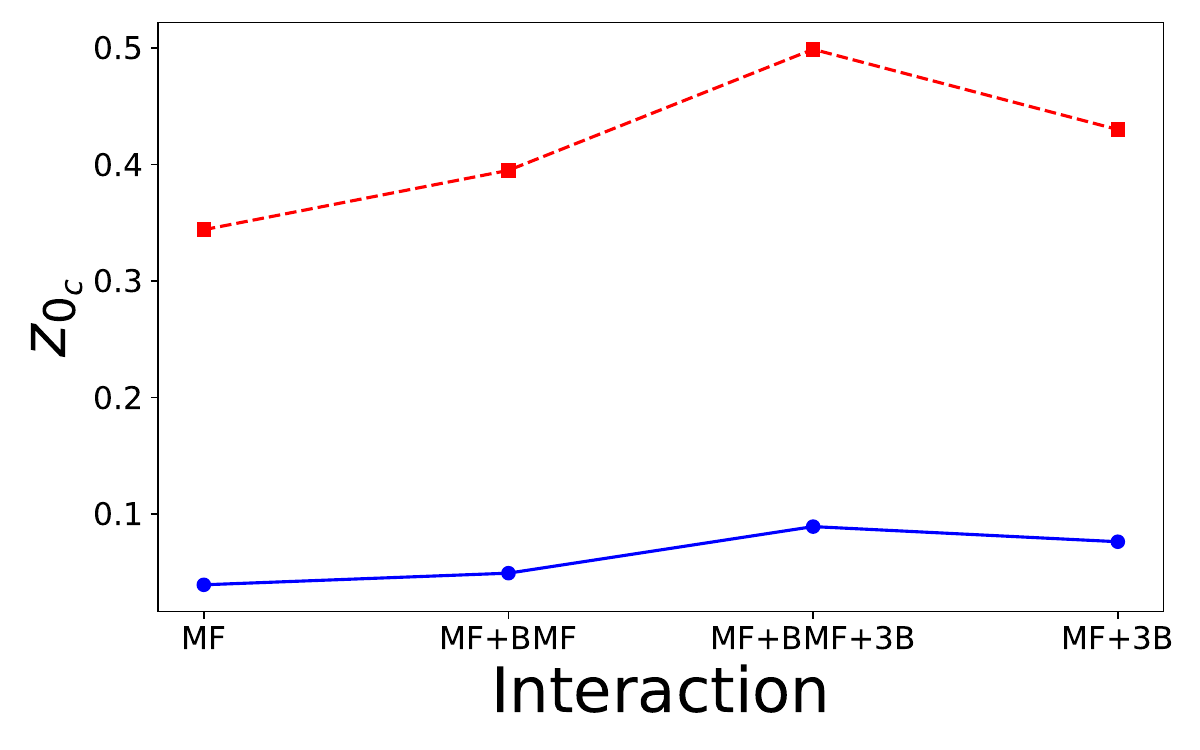}
	\caption{(Color Online) Variation of critical imbalance for different interaction competition in SDW and ADW potential in Q1D system. The blue solid line describes the symmetric case while the red dashed line describes the asymmetric double well problem.}\label{sdw_adw}
	\end{figure}

One noticeable difference from symmetric to asymmetric case is that the oscillation frequency is actually gets suppressed when there is more competition between different interactions.    
It also must be noted that, the variation of frequency is relatively small in the symmetric cases while a small asymmetry allows relatively significant variation in the frequency. As a result, we expect experimental verification of BMF contribution via Josephson effect for a tilted lattice can be observed. We can clearly identify this from Fig.~\ref{sdw_adw} where the critical imbalance is noted for various interaction competition in SDW and ADW. It suggests the ST regime is significantly more distinct and experimentally accessible in the asymmetric case compared to the extremely low thresholds found in the symmetric Q1D case.

In Table.\ref{Table2}, we have noted the critical imbalance beyond which we observe the ST phase. In Q1D system, (for both symmetric and asymmetric potential), the critical imbalance increases as the system is subjected to more interaction competition which it is reversed in symmetric 1D system (even though the change is marginal). Here, it must also be noted that the complementary nature of MF and BMF interaction in Q1D and 1D system is directly responsible for this feature. 
\begin{table}
	\centering
	\caption{Observing critical imbalance value for  different permutation of interactions}\label{Table2}
\vskip 0.2cm
	\begin{tabular}{c@{\hskip 1pt} c@{\hskip 1pt} c@{\hskip 3pt} c@{\hskip 3pt} c@{\hskip 1pt} c }
		\hline \hline
		Interaction  & $ (z_{0_{c}(Q1D)})_{SDW}$ & $(z_{0_{c}(1D)})_{SDW} $ & $ (z_{0_{c}(Q1D)})_{ADW} $\\  
		\\ \hline \\
		MF & 0.039 & 0.285& 0.344  \\ [0.3cm]
		
		MF + BMF  & 0.049 & 0.284 & 0.395  \\[0.3cm]
		
		MF + BMF + 3B & 0.089& 0.275 & 0.499\\ [0.3cm]
		
		MF + 3B & 0.076& 0.276& 0.43 \\
		\hline \hline
	\end{tabular}
\end{table}

\section{Bogoliubov Method}\label{quasi-particle}
The physics of double-well potential is well captured via TMM, nevertheless a complimentary description by means of Bogoliubov quasiparticle method is considered to be more educative \cite{P.wysocki_josephson_2024}. One can actually provide a direct correspondence between Bogoliubov description and two mode approximation in alkali atomic gases \cite{paraoanu2001josephson}. It has been reported that for a very small imbalance between the left and right wells Bogoliubov modes with lowest energy is occupied. This is the regime where $\omega_J\sim\omega_B$ as we observe a periodic transfer of population between the ground and the first excited state. When the initial imbalance is increased, more excited modes get coupled during the evolution of the system. This nonlinear mixing between Bogoliubov modes eventually drives the system into the the self-trapping regime, when the population of the ground state can be transferred completely to the excited states \cite{sofia}.

More precisely, one can comment that at relatively weak interaction and low momentum, the Josephson and Bogoliubov frequencies do match quite appreciably \cite{burchianti2017josephson}. In repulsive BEC, the Bogoliubov frequency in general remains real as $\omega_B^2\sim J(J+\Lambda)$. This allows a smooth transition from JO to ST which is already been reported \cite{smerzi, raghavan1999coherent} and later have been extended for BMF effect \cite{P.wysocki_josephson_2024}. 

We must reiterate here that, in 1D the mean-field interaction is repulsive while the BMF interaction (which is relatively weak in nature) is attractive. In Q1D system, the situation is little different as MF interaction is attractive and BMF interaction is repulsive. We have seen in our calculation that $\Lambda$ is negative which can lead to a domain where $\omega_B$ is complex resulting dynamical instability. Therefore study of Bogoliubov spectrum becomes more important.


Here we carry out the quasi particle excitation calculation for a quantum gas which is subjected to BMF and 3B effects in addition to the usual MF interaction. Through our analysis we obtain the Bogoliubov frequency which agrees well with the Josephson frequency in the low momenta region. Additionally, we note roton like dispersion minima at finite momentum and low density. 

We start by assuming a small perturbation of $\delta\psi(x,t)$ on the ground state such that $\psi(x,t)=e^{-i\mu t} (\psi_0(x)+\delta\psi(x,t))$. Applying this ansatz in Eq.(\ref{GPE:BMF}) and linearizing for $\delta\psi$ we can obtain a set of equation for Q1D and 1D respectively. Next, expanding the perturbation into the quasiparticle modes as, $\delta\psi(x,t)=e^{-i\mu t}  \sum_q\left[u_q(x)e^{-i\omega t}-v_q^{*}(x)e^{i\omega t}\right]$ it is possible to obtain an equation for $z(t)$ as a function of Bogoliubov frequency ($\omega$) which highlights the good agreement between Josephson frequency and Bogoliubov frequency \cite{burchianti2017josephson}. The Bogoliubov de-Gennes (BdG) equation for Q1D and 1D system can now be noted as,
\begin{equation}
\begin{pmatrix}
\mathcal{K}_d-\omega & \mathcal{M}_d\\
\mathcal{M}_d & \mathcal{K}_d+\omega,
\end{pmatrix}\begin{pmatrix}u_q\\v_q\end{pmatrix}=0,
\end{equation}
where $d\in\lbrace Q1D, 1D\rbrace$, $\mathcal{K}_{Q1D}=\frac{1}{2}q^2+2g_3|\psi_0|^2 +\frac{5}{2}g_4 |\psi_0|^3 + 3g_5 |\psi_0|^4 - \mu$, $\mathcal{K}_{1D}=\frac{1}{2}q^2+2g_3|\psi_0|^2 \frac{3}{2}g_2 |\psi_0| + 3g_5 |\psi_0|^4 - \mu$, $\mathcal{M}_{Q1D}=g_3|\psi_0|^2 +\frac{3}{2}g_4|\psi_0|^3 + 2g_5 |\psi_0|^4$ and $\mathcal{M}_{1D}=g_3|\psi_0|^2 +\frac{1}{2}g_2|\psi_0|+2g_5 |\psi_0|^4$. 

We know that for unique solution of $u_q$ and $v_q$ the determinant of the BdG matrix should be equal to zero. This allows us to write down the dispersion equations for both the dimensions in the  following way: 
\begin{eqnarray}\label{dispersion}
	\omega_{Q1D}&=&\sqrt{\frac{q^4}{4}+\mathcal{A}q^2+\mathcal{B}}\\
	\omega_{1D}&=&\sqrt{\frac{q^4}{4} + \mathcal{C}q^2 +\mathcal{D}},\\
	\mathcal{A}&=&\frac{5}{2}|\psi_0|^3g_4+3|\psi_0|^4g_5+2|\psi_0|^2g_3-\mu,\nonumber\\
	\mathcal{B}&=&7|\psi_0|^5g_3g_4+8|\psi_0|^6g_3g_5+4|\psi_0|^6g_4^2\nonumber\\&&+9|\psi_0|^7g_4g_5+5|\psi_0|^8g_5^2+3|\psi_0|^4g_3^2+\mu^2\nonumber\\&&-\mu\left(4|\psi_0|^2g_3+5|\psi_0|^3g_4+6|\psi_0|^4g_5\right)\nonumber\\
	\mathcal{C}&=&\frac{3}{2}g_2|\psi_0| + 3 |\psi_0|^4g_5 +2 |\psi_0|^2 g_3-\mu,\nonumber\\
	\mathcal{D}&=&5|\psi_0|^3 g_2g_3 + 8|\psi_0|^6g_3g_5 + 2|\psi_0|^2g_2^2\nonumber\\&&+7 |\psi_0|^5g_2g_5+5|\psi_0|^8g_5^2+3|\psi_0|^4g_3^2+\mu^2\nonumber\\&&-\mu\left(4|\psi_0|^2g_3+3|\psi_0|g_2+6|\psi_0|^4g_5\right)
\end{eqnarray}

From Eq.(\ref{dispersion}) we can clearly see that in absence of BMF and 3B interactions, our dispersion relation boils down to the well known Bogoliubov dispersion with Thomas-Fermi approximation ($\omega_B=\sqrt{\frac{q^4}{4}+g_3|\psi_0|^2q^2}$). Here, we have not assigned any sign to $g_3$, $g_2$ or $g_4$, however, it must be noted that in Q1D system $g_3$ is attractive and the BMF contribution, $g_4$ is repulsive while in 1D the situation is reversed, implying $g_3$ as repulsive and $g_2$ as attractive. We have considered $g_5>0$ however, we did not note any significant change in result for $g_5<0$. This is primarily due to weak three-body interaction strength. Contrary to Ref.~\cite{P.wysocki_josephson_2024}, we have used the BMF interaction strength as about one order of magnitude less, instead of same order of magnitude as considered therein. In the same spirit we have used 3B interaction strength also considerably small when compared with MF interaction.

\begin{figure}
	\centering
	\includegraphics[width = 1.0\linewidth]{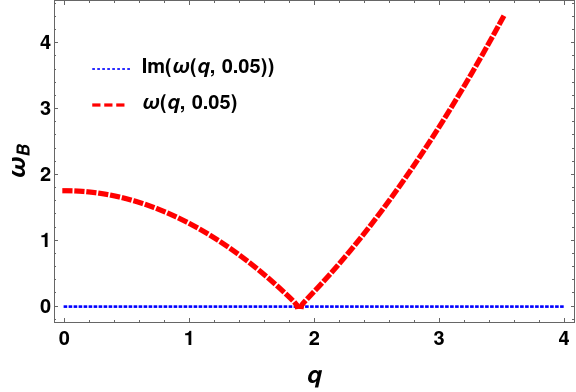}
	\caption{Bogoliubov dispersion in Q1D system. $g_3=-0.6$, $g_4=0.01$, $g_5=0.009$ and $\mu=1.7$. The critical momentum corresponding to the dispersion kink is noted as $q_c=1.84$.}\label{q1d_dis}
\end{figure}
\begin{figure}
	\centering
	\includegraphics[width = 1.0\linewidth]{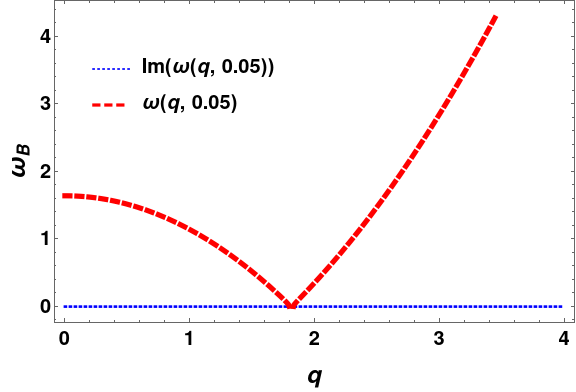}
	\caption{Bogoliubov dispersion in 1D system. $g_3=0.6$, $g_2=-0.01$, $g_5=0.009$ and $\mu=1.7$. The critical momentum corresponding to the dispersion kink is noted as $q_c=1.86$.}\label{1d_dis}
\end{figure}

We have depicted the dispersion in both Q1D and 1D system. Interestingly, the interplay of different interaction components leads to a nontrivial dispersion as displayed in Fig.~\ref{q1d_dis} and ~\ref{1d_dis}. Using the numerical values from previous section for all the interaction parameters and the numerically obtained energy difference between two modes as the seed value of chemical potential, we observe roton-like mode at very low density (or low normalization). The figures also reveal that at low density our system is dynamically stable as $\mathrm{Im}(\omega_B)=0$.

Further we observe the critical momentum where $\omega_B\rightarrow 0$ is almost same for both Q1D and 1D systems. It is noted as $q\simeq 1.84$ in Q1D and $1.86$ for 1D. The red-dashed line describes the Bogolioubov frequency while the blue-dotted line corresponds to the complex part of Bogoliubov frequency. 


The finite $\omega_B$ for $q\rightarrow 0$ describes the roton energy gap which depends on the density and on the strength of the interactions \cite{chomaz}. The dispersion relation also enlightens us about the region of instability where $\omega_B$ becomes complex quantity. Very recently it is been shown that the roton instability can lead to the generation of super-solid like phase in dipolar condensate \cite{alana2023supersolid}. Nonetheless, we must categorically add here that the system under consideration does not take into account the dipolar interaction. The `kink' in the dispersion is solely arises from the interaction competitions. The dispersion calculation for BEC in the purview of GP equation is well known ($\omega=\sqrt{q^4/4+g_3|\psi_0|^2q^2}$, as mentioned earlier). To obtain the dispersion relation in BEC we also apply the well known Thomas-Fermi (TF) approximation where it is assumed that $\mu=gn$. However, the current description enlightens us to a more complex effective interaction landscape. A deeper inspection of Eq.(\ref{dispersion}) reveals that the origin of the `kink' is existence of non-zero $\mathcal{A, B, C}$ and $\mathcal{D}$ for Q1D and 1D respectively. Based on the expressions provided in Eq.(\ref{dispersion}), it is possible to adjust the interactions such a way that $\mathcal{B}$ (in Q1D) or $\mathcal{D}$ (in 1D) are zero. In those cases the dispersion relation provides the domain of modulational instability with side-lobe generation from the imaginary part of the frequency. We have recently reported this situation for a dipolar condensate \cite{singh2026fragmentation}. Due to fierce competition between the interactions (see Eq.(\ref{dispersion})), it is also possible that $\mathcal{A}$ or $\mathcal{C}$ is zero. That leads to the well known parabolic dispersion spectrum. Even at the regime of low wave-number one can obtain a non-trivial sound mode. It must be reminded here that in regular BEC (in the purview of GP equation), we obtain sound velocity as $\omega=c_sq$, $c_s$ being the sound velocity, when $q\rightarrow0$,  Albeit, all these interesting prospects demand careful analysis, we limit ourselves in the observation of Roton-like modes at this juncture and we plan to revisit it in our forthcoming investigations. Hence, in a nut-shell, we hope the current discussion can be matter of deeper investigation in the near future.

\section{Conclusion}\label{conclusion}
In conclusion, we would like to summarize that we have studied particle excitation in Bose system in quasi one dimension and also noted the complementary results in one dimension when the particles are trapped in a double well potential and experiences different interaction competitions. Primarily we follow the two-mode model however, we also incorporate the Bogoliubov quasiparticle method for additional insights. We have further extended our analysis to asymmetric double well potential as this tilted topology is key part of atomtronics where it enables the study of coherent quantum transport and the creation of atom-based SQUIDs \cite{borysenko2025acceleration}.

Even though, JO and ST is a well studied problem in repulsive BEC however, in attractive BEC it is never been analyzed earlier. This is due to the fact that for observing ST regime the scaled Hamiltonian should be greater than $1$, while in attractive BEC it is less than zero at $0$ relative phase. Here, we show JO to ST like transition in an attractive BEC when the relative phase is fixed at $\pi$. We further note that, it is even possible to observe ST regime for a very weak attractive mean-field interaction and a very strong repulsive BMF interaction as $\omega_B\sim\sqrt{J(J+\Lambda)}\in\mathbb{R}$. Through Bogoloiubov spectrum we also note that a dynamically stable JO to ST transition is possible at low density \cite{P.wysocki_josephson_2024}. 

Further we note that, the Josephson frequency is non-monotonic as a function of particle number in Q1D (just like earlier reported for 1D), however, for relatively large particle number three-body effect dominates and straitens the Josephson frequency. We have also systematically noted the Josephson frequency at a fixed imbalance for different interaction competition and realize that addition of small asymmetry actually pronounces the interaction competition effect. This can lead towards observation of Josephson frequency for different interaction competition experimentally and can enrich our understanding more towards quantum many body effect. We have also carefully noted the variation of critical imbalance beyond which one can observe a transition between Josephson regime and self trapping and presented a phase diagram describing the same. The critical imbalance is lowest in quasi one dimension when the particles are trapped in a symmetric double well potential, while it is relatively higher in asymmetric double well potential enabling us to conclude that the ST regime is experimentally more accessible in the asymmetric case.

Our Bogoliubov excitation calculation qualitatively corroborates with our TMM result but importantly it allows us to identify an exiting roton-like mode as well. We note the critical momentum where a frequency kink is present. Although, roton-like mode is experimentally observed and noted as a signature of the onset of super-solid like phase \cite{chomaz, roccuzzo2019supersolid}, nevertheless the observed dispersion `kink' here can be matter of interest and further deeper investigation and can lead to appropriate characterization of this behavior. 

We hope that the current analysis will be useful for future experimental investigation on the role of interaction, dimensionality, and asymmetry. 
\section*{Declaration} 
\subsection*{Funding}
AK thanks Anusandhan National Research Foundation (ANRF), Department of Science and Technology (DST), India for the support provided through the project number CRG/2023/001220. 
\subsection*{Availability of Data and Material}
Data and Material will be made available on request.
\subsection*{Competing Interests}
There is no competing interests involved for any of the authors.
\subsection*{Authors Contributions}
S.S.: TMM Calculation, Visualization, Interpretation, Writing, Reviewing and Editing; Ibrar: Bogoliubov Calculation, Visualization, Interpretation; S.R.: Initiation of Bogoliubov Calculation; A.K.: Conceptualization, Formal analysis and Interpretation, Writing – Original Draft, Writing – Review \& Editing, Visualization. 
\bibliographystyle{apsrev4-1}
\bibliography{ref_M1}

\begin{thebibliography}{57}%
\makeatletter
\providecommand \@ifxundefined [1]{%
 \@ifx{#1\undefined}
}%
\providecommand \@ifnum [1]{%
 \ifnum #1\expandafter \@firstoftwo
 \else \expandafter \@secondoftwo
 \fi
}%
\providecommand \@ifx [1]{%
 \ifx #1\expandafter \@firstoftwo
 \else \expandafter \@secondoftwo
 \fi
}%
\providecommand \natexlab [1]{#1}%
\providecommand \enquote  [1]{``#1''}%
\providecommand \bibnamefont  [1]{#1}%
\providecommand \bibfnamefont [1]{#1}%
\providecommand \citenamefont [1]{#1}%
\providecommand \href@noop [0]{\@secondoftwo}%
\providecommand \href [0]{\begingroup \@sanitize@url \@href}%
\providecommand \@href[1]{\@@startlink{#1}\@@href}%
\providecommand \@@href[1]{\endgroup#1\@@endlink}%
\providecommand \@sanitize@url [0]{\catcode `\\12\catcode `\$12\catcode
  `\&12\catcode `\#12\catcode `\^12\catcode `\_12\catcode `\%12\relax}%
\providecommand \@@startlink[1]{}%
\providecommand \@@endlink[0]{}%
\providecommand \url  [0]{\begingroup\@sanitize@url \@url }%
\providecommand \@url [1]{\endgroup\@href {#1}{\urlprefix }}%
\providecommand \urlprefix  [0]{URL }%
\providecommand \Eprint [0]{\href }%
\providecommand \doibase [0]{http://dx.doi.org/}%
\providecommand \selectlanguage [0]{\@gobble}%
\providecommand \bibinfo  [0]{\@secondoftwo}%
\providecommand \bibfield  [0]{\@secondoftwo}%
\providecommand \translation [1]{[#1]}%
\providecommand \BibitemOpen [0]{}%
\providecommand \bibitemStop [0]{}%
\providecommand \bibitemNoStop [0]{.\EOS\space}%
\providecommand \EOS [0]{\spacefactor3000\relax}%
\providecommand \BibitemShut  [1]{\csname bibitem#1\endcsname}%
\let\auto@bib@innerbib\@empty
\bibitem [{\citenamefont {Devoret}\ \emph {et~al.}(1985)\citenamefont
  {Devoret}, \citenamefont {Martinis},\ and\ \citenamefont {Clarke}}]{devoret}%
  \BibitemOpen
  \bibfield  {author} {\bibinfo {author} {\bibfnamefont {M.~H.}\ \bibnamefont
  {Devoret}}, \bibinfo {author} {\bibfnamefont {J.~M.}\ \bibnamefont
  {Martinis}}, \ and\ \bibinfo {author} {\bibfnamefont {J.}~\bibnamefont
  {Clarke}},\ }\href@noop {} {\bibfield  {journal} {\bibinfo  {journal}
  {Physical review letters}\ }\textbf {\bibinfo {volume} {55}},\ \bibinfo
  {pages} {1908} (\bibinfo {year} {1985})}\BibitemShut {NoStop}%
\bibitem [{\citenamefont {You}\ and\ \citenamefont {Nori}(2005)}]{nori}%
  \BibitemOpen
  \bibfield  {author} {\bibinfo {author} {\bibfnamefont {J.}~\bibnamefont
  {You}}\ and\ \bibinfo {author} {\bibfnamefont {F.}~\bibnamefont {Nori}},\
  }\href@noop {} {\bibfield  {journal} {\bibinfo  {journal} {Physics today}\
  }\textbf {\bibinfo {volume} {58}},\ \bibinfo {pages} {42} (\bibinfo {year}
  {2005})}\BibitemShut {NoStop}%
\bibitem [{\citenamefont {Yu}\ \emph {et~al.}(2002)\citenamefont {Yu},
  \citenamefont {Han}, \citenamefont {Chu}, \citenamefont {Chu},\ and\
  \citenamefont {Wang}}]{yu}%
  \BibitemOpen
  \bibfield  {author} {\bibinfo {author} {\bibfnamefont {Y.}~\bibnamefont
  {Yu}}, \bibinfo {author} {\bibfnamefont {S.}~\bibnamefont {Han}}, \bibinfo
  {author} {\bibfnamefont {X.}~\bibnamefont {Chu}}, \bibinfo {author}
  {\bibfnamefont {S.-I.}\ \bibnamefont {Chu}}, \ and\ \bibinfo {author}
  {\bibfnamefont {Z.}~\bibnamefont {Wang}},\ }\href@noop {} {\bibfield
  {journal} {\bibinfo  {journal} {Science}\ }\textbf {\bibinfo {volume}
  {296}},\ \bibinfo {pages} {889} (\bibinfo {year} {2002})}\BibitemShut
  {NoStop}%
\bibitem [{\citenamefont {Martinis}\ \emph {et~al.}(2002)\citenamefont
  {Martinis}, \citenamefont {Nam}, \citenamefont {Aumentado},\ and\
  \citenamefont {Urbina}}]{martinis}%
  \BibitemOpen
  \bibfield  {author} {\bibinfo {author} {\bibfnamefont {J.~M.}\ \bibnamefont
  {Martinis}}, \bibinfo {author} {\bibfnamefont {S.}~\bibnamefont {Nam}},
  \bibinfo {author} {\bibfnamefont {J.}~\bibnamefont {Aumentado}}, \ and\
  \bibinfo {author} {\bibfnamefont {C.}~\bibnamefont {Urbina}},\ }\href@noop {}
  {\bibfield  {journal} {\bibinfo  {journal} {Physical review letters}\
  }\textbf {\bibinfo {volume} {89}},\ \bibinfo {pages} {117901} (\bibinfo
  {year} {2002})}\BibitemShut {NoStop}%
\bibitem [{\citenamefont {Cazalilla}\ \emph {et~al.}(2011)\citenamefont
  {Cazalilla}, \citenamefont {Citro}, \citenamefont {Giamarchi}, \citenamefont
  {Orignac},\ and\ \citenamefont {Rigol}}]{cazalilla2011one}%
  \BibitemOpen
  \bibfield  {author} {\bibinfo {author} {\bibfnamefont {M.~A.}\ \bibnamefont
  {Cazalilla}}, \bibinfo {author} {\bibfnamefont {R.}~\bibnamefont {Citro}},
  \bibinfo {author} {\bibfnamefont {T.}~\bibnamefont {Giamarchi}}, \bibinfo
  {author} {\bibfnamefont {E.}~\bibnamefont {Orignac}}, \ and\ \bibinfo
  {author} {\bibfnamefont {M.}~\bibnamefont {Rigol}},\ }\href@noop {}
  {\bibfield  {journal} {\bibinfo  {journal} {Reviews of Modern Physics}\
  }\textbf {\bibinfo {volume} {83}},\ \bibinfo {pages} {1405} (\bibinfo {year}
  {2011})}\BibitemShut {NoStop}%
\bibitem [{\citenamefont {Dutta}\ \emph {et~al.}(2015)\citenamefont {Dutta},
  \citenamefont {Barman}, \citenamefont {Siddharth}, \citenamefont {Khan},\
  and\ \citenamefont {Basu}}]{dutta2015tunneling}%
  \BibitemOpen
  \bibfield  {author} {\bibinfo {author} {\bibfnamefont {S.}~\bibnamefont
  {Dutta}}, \bibinfo {author} {\bibfnamefont {A.}~\bibnamefont {Barman}},
  \bibinfo {author} {\bibfnamefont {A.}~\bibnamefont {Siddharth}}, \bibinfo
  {author} {\bibfnamefont {A.}~\bibnamefont {Khan}}, \ and\ \bibinfo {author}
  {\bibfnamefont {S.}~\bibnamefont {Basu}},\ }\href@noop {} {\bibfield
  {journal} {\bibinfo  {journal} {The European Physical Journal B}\ }\textbf
  {\bibinfo {volume} {88}},\ \bibinfo {pages} {139} (\bibinfo {year}
  {2015})}\BibitemShut {NoStop}%
\bibitem [{\citenamefont {Ngo}\ \emph {et~al.}(2021)\citenamefont {Ngo},
  \citenamefont {Tsarev}, \citenamefont {Lee},\ and\ \citenamefont
  {Alodjants}}]{nature_Qmetro_2021BEC}%
  \BibitemOpen
  \bibfield  {author} {\bibinfo {author} {\bibfnamefont {T.~V.}\ \bibnamefont
  {Ngo}}, \bibinfo {author} {\bibfnamefont {D.~V.}\ \bibnamefont {Tsarev}},
  \bibinfo {author} {\bibfnamefont {R.-K.}\ \bibnamefont {Lee}}, \ and\
  \bibinfo {author} {\bibfnamefont {A.~P.}\ \bibnamefont {Alodjants}},\
  }\href@noop {} {\bibfield  {journal} {\bibinfo  {journal} {Scientific
  Reports}\ }\textbf {\bibinfo {volume} {11}},\ \bibinfo {pages} {19363}
  (\bibinfo {year} {2021})}\BibitemShut {NoStop}%
\bibitem [{\citenamefont {Amico}\ \emph {et~al.}(2021)\citenamefont {Amico},
  \citenamefont {Boshier}, \citenamefont {Birkl}, \citenamefont {Minguzzi},
  \citenamefont {Miniatura}, \citenamefont {Kwek}, \citenamefont {Aghamalyan},
  \citenamefont {Ahufinger}, \citenamefont {Anderson}, \citenamefont {Andrei}
  \emph {et~al.}}]{amico}%
  \BibitemOpen
  \bibfield  {author} {\bibinfo {author} {\bibfnamefont {L.}~\bibnamefont
  {Amico}}, \bibinfo {author} {\bibfnamefont {M.}~\bibnamefont {Boshier}},
  \bibinfo {author} {\bibfnamefont {G.}~\bibnamefont {Birkl}}, \bibinfo
  {author} {\bibfnamefont {A.}~\bibnamefont {Minguzzi}}, \bibinfo {author}
  {\bibfnamefont {C.}~\bibnamefont {Miniatura}}, \bibinfo {author}
  {\bibfnamefont {L.-C.}\ \bibnamefont {Kwek}}, \bibinfo {author}
  {\bibfnamefont {D.}~\bibnamefont {Aghamalyan}}, \bibinfo {author}
  {\bibfnamefont {V.}~\bibnamefont {Ahufinger}}, \bibinfo {author}
  {\bibfnamefont {D.}~\bibnamefont {Anderson}}, \bibinfo {author}
  {\bibfnamefont {N.}~\bibnamefont {Andrei}},  \emph {et~al.},\ }\href@noop {}
  {\bibfield  {journal} {\bibinfo  {journal} {AVS Quantum Science}\ }\textbf
  {\bibinfo {volume} {3}} (\bibinfo {year} {2021})}\BibitemShut {NoStop}%
\bibitem [{\citenamefont {Degen}\ \emph {et~al.}(2017)\citenamefont {Degen},
  \citenamefont {Reinhard},\ and\ \citenamefont
  {Cappellaro}}]{degen_2017quantum}%
  \BibitemOpen
  \bibfield  {author} {\bibinfo {author} {\bibfnamefont {C.~L.}\ \bibnamefont
  {Degen}}, \bibinfo {author} {\bibfnamefont {F.}~\bibnamefont {Reinhard}}, \
  and\ \bibinfo {author} {\bibfnamefont {P.}~\bibnamefont {Cappellaro}},\
  }\href@noop {} {\bibfield  {journal} {\bibinfo  {journal} {Reviews of modern
  physics}\ }\textbf {\bibinfo {volume} {89}},\ \bibinfo {pages} {035002}
  (\bibinfo {year} {2017})}\BibitemShut {NoStop}%
\bibitem [{\citenamefont {Sch{\"a}fer}\ \emph {et~al.}(2020)\citenamefont
  {Sch{\"a}fer}, \citenamefont {Fukuhara}, \citenamefont {Sugawa},
  \citenamefont {Takasu},\ and\ \citenamefont {Takahashi}}]{schafer}%
  \BibitemOpen
  \bibfield  {author} {\bibinfo {author} {\bibfnamefont {F.}~\bibnamefont
  {Sch{\"a}fer}}, \bibinfo {author} {\bibfnamefont {T.}~\bibnamefont
  {Fukuhara}}, \bibinfo {author} {\bibfnamefont {S.}~\bibnamefont {Sugawa}},
  \bibinfo {author} {\bibfnamefont {Y.}~\bibnamefont {Takasu}}, \ and\ \bibinfo
  {author} {\bibfnamefont {Y.}~\bibnamefont {Takahashi}},\ }\href@noop {}
  {\bibfield  {journal} {\bibinfo  {journal} {Nature Reviews Physics}\ }\textbf
  {\bibinfo {volume} {2}},\ \bibinfo {pages} {411} (\bibinfo {year}
  {2020})}\BibitemShut {NoStop}%
\bibitem [{\citenamefont {Smerzi}\ \emph {et~al.}(1997)\citenamefont {Smerzi},
  \citenamefont {Fantoni}, \citenamefont {Giovanazzi},\ and\ \citenamefont
  {Shenoy}}]{smerzi}%
  \BibitemOpen
  \bibfield  {author} {\bibinfo {author} {\bibfnamefont {A.}~\bibnamefont
  {Smerzi}}, \bibinfo {author} {\bibfnamefont {S.}~\bibnamefont {Fantoni}},
  \bibinfo {author} {\bibfnamefont {S.}~\bibnamefont {Giovanazzi}}, \ and\
  \bibinfo {author} {\bibfnamefont {S.}~\bibnamefont {Shenoy}},\ }\href@noop {}
  {\bibfield  {journal} {\bibinfo  {journal} {Physical Review Letters}\
  }\textbf {\bibinfo {volume} {79}},\ \bibinfo {pages} {4950} (\bibinfo {year}
  {1997})}\BibitemShut {NoStop}%
\bibitem [{\citenamefont {Albiez}\ \emph {et~al.}(2005)\citenamefont {Albiez},
  \citenamefont {Gati}, \citenamefont {F{\"o}lling}, \citenamefont {Hunsmann},
  \citenamefont {Cristiani},\ and\ \citenamefont {Oberthaler}}]{gati}%
  \BibitemOpen
  \bibfield  {author} {\bibinfo {author} {\bibfnamefont {M.}~\bibnamefont
  {Albiez}}, \bibinfo {author} {\bibfnamefont {R.}~\bibnamefont {Gati}},
  \bibinfo {author} {\bibfnamefont {J.}~\bibnamefont {F{\"o}lling}}, \bibinfo
  {author} {\bibfnamefont {S.}~\bibnamefont {Hunsmann}}, \bibinfo {author}
  {\bibfnamefont {M.}~\bibnamefont {Cristiani}}, \ and\ \bibinfo {author}
  {\bibfnamefont {M.~K.}\ \bibnamefont {Oberthaler}},\ }\href@noop {}
  {\bibfield  {journal} {\bibinfo  {journal} {Physical review letters}\
  }\textbf {\bibinfo {volume} {95}},\ \bibinfo {pages} {010402} (\bibinfo
  {year} {2005})}\BibitemShut {NoStop}%
\bibitem [{\citenamefont {Mart{\'\i}nez-Garaot}\ \emph
  {et~al.}(2018)\citenamefont {Mart{\'\i}nez-Garaot}, \citenamefont {Pettini},\
  and\ \citenamefont {Modugno}}]{sofia}%
  \BibitemOpen
  \bibfield  {author} {\bibinfo {author} {\bibfnamefont {S.}~\bibnamefont
  {Mart{\'\i}nez-Garaot}}, \bibinfo {author} {\bibfnamefont {G.}~\bibnamefont
  {Pettini}}, \ and\ \bibinfo {author} {\bibfnamefont {M.}~\bibnamefont
  {Modugno}},\ }\href@noop {} {\bibfield  {journal} {\bibinfo  {journal}
  {Physical Review A}\ }\textbf {\bibinfo {volume} {98}},\ \bibinfo {pages}
  {043624} (\bibinfo {year} {2018})}\BibitemShut {NoStop}%
\bibitem [{\citenamefont {Vivek}\ \emph {et~al.}(2025)\citenamefont {Vivek},
  \citenamefont {Mondal}, \citenamefont {Chakraborty},\ and\ \citenamefont
  {Sinha}}]{vivek}%
  \BibitemOpen
  \bibfield  {author} {\bibinfo {author} {\bibfnamefont {G.}~\bibnamefont
  {Vivek}}, \bibinfo {author} {\bibfnamefont {D.}~\bibnamefont {Mondal}},
  \bibinfo {author} {\bibfnamefont {S.}~\bibnamefont {Chakraborty}}, \ and\
  \bibinfo {author} {\bibfnamefont {S.}~\bibnamefont {Sinha}},\ }\href@noop {}
  {\bibfield  {journal} {\bibinfo  {journal} {Physical Review Letters}\
  }\textbf {\bibinfo {volume} {134}},\ \bibinfo {pages} {113404} (\bibinfo
  {year} {2025})}\BibitemShut {NoStop}%
\bibitem [{\citenamefont {Kadau}\ \emph {et~al.}(2016)\citenamefont {Kadau},
  \citenamefont {Schmitt}, \citenamefont {Wenzel}, \citenamefont {Wink},
  \citenamefont {Maier}, \citenamefont {Ferrier-Barbut},\ and\ \citenamefont
  {Pfau}}]{kadau}%
  \BibitemOpen
  \bibfield  {author} {\bibinfo {author} {\bibfnamefont {H.}~\bibnamefont
  {Kadau}}, \bibinfo {author} {\bibfnamefont {M.}~\bibnamefont {Schmitt}},
  \bibinfo {author} {\bibfnamefont {M.}~\bibnamefont {Wenzel}}, \bibinfo
  {author} {\bibfnamefont {C.}~\bibnamefont {Wink}}, \bibinfo {author}
  {\bibfnamefont {T.}~\bibnamefont {Maier}}, \bibinfo {author} {\bibfnamefont
  {I.}~\bibnamefont {Ferrier-Barbut}}, \ and\ \bibinfo {author} {\bibfnamefont
  {T.}~\bibnamefont {Pfau}},\ }\href@noop {} {\bibfield  {journal} {\bibinfo
  {journal} {Nature}\ }\textbf {\bibinfo {volume} {530}},\ \bibinfo {pages}
  {194} (\bibinfo {year} {2016})}\BibitemShut {NoStop}%
\bibitem [{\citenamefont {Ferrier-Barbut}\ \emph {et~al.}(2016)\citenamefont
  {Ferrier-Barbut}, \citenamefont {Kadau}, \citenamefont {Schmitt},
  \citenamefont {Wenzel},\ and\ \citenamefont {Pfau}}]{pfau}%
  \BibitemOpen
  \bibfield  {author} {\bibinfo {author} {\bibfnamefont {I.}~\bibnamefont
  {Ferrier-Barbut}}, \bibinfo {author} {\bibfnamefont {H.}~\bibnamefont
  {Kadau}}, \bibinfo {author} {\bibfnamefont {M.}~\bibnamefont {Schmitt}},
  \bibinfo {author} {\bibfnamefont {M.}~\bibnamefont {Wenzel}}, \ and\ \bibinfo
  {author} {\bibfnamefont {T.}~\bibnamefont {Pfau}},\ }\href@noop {} {\bibfield
   {journal} {\bibinfo  {journal} {Physical review letters}\ }\textbf {\bibinfo
  {volume} {116}},\ \bibinfo {pages} {215301} (\bibinfo {year}
  {2016})}\BibitemShut {NoStop}%
\bibitem [{\citenamefont {Cabrera}\ \emph {et~al.}(2018)\citenamefont
  {Cabrera}, \citenamefont {Tanzi}, \citenamefont {Sanz}, \citenamefont
  {Naylor}, \citenamefont {Thomas}, \citenamefont {Cheiney},\ and\
  \citenamefont {Tarruell}}]{cab}%
  \BibitemOpen
  \bibfield  {author} {\bibinfo {author} {\bibfnamefont {C.}~\bibnamefont
  {Cabrera}}, \bibinfo {author} {\bibfnamefont {L.}~\bibnamefont {Tanzi}},
  \bibinfo {author} {\bibfnamefont {J.}~\bibnamefont {Sanz}}, \bibinfo {author}
  {\bibfnamefont {B.}~\bibnamefont {Naylor}}, \bibinfo {author} {\bibfnamefont
  {P.}~\bibnamefont {Thomas}}, \bibinfo {author} {\bibfnamefont
  {P.}~\bibnamefont {Cheiney}}, \ and\ \bibinfo {author} {\bibfnamefont
  {L.}~\bibnamefont {Tarruell}},\ }\href@noop {} {\bibfield  {journal}
  {\bibinfo  {journal} {Science}\ }\textbf {\bibinfo {volume} {359}},\ \bibinfo
  {pages} {301} (\bibinfo {year} {2018})}\BibitemShut {NoStop}%
\bibitem [{\citenamefont {Cheiney}\ \emph {et~al.}(2018)\citenamefont
  {Cheiney}, \citenamefont {Cabrera}, \citenamefont {Sanz}, \citenamefont
  {Naylor}, \citenamefont {Tanzi},\ and\ \citenamefont {Tarruell}}]{che}%
  \BibitemOpen
  \bibfield  {author} {\bibinfo {author} {\bibfnamefont {P.}~\bibnamefont
  {Cheiney}}, \bibinfo {author} {\bibfnamefont {C.}~\bibnamefont {Cabrera}},
  \bibinfo {author} {\bibfnamefont {J.}~\bibnamefont {Sanz}}, \bibinfo {author}
  {\bibfnamefont {B.}~\bibnamefont {Naylor}}, \bibinfo {author} {\bibfnamefont
  {L.}~\bibnamefont {Tanzi}}, \ and\ \bibinfo {author} {\bibfnamefont
  {L.}~\bibnamefont {Tarruell}},\ }\href@noop {} {\bibfield  {journal}
  {\bibinfo  {journal} {Physical review letters}\ }\textbf {\bibinfo {volume}
  {120}},\ \bibinfo {pages} {135301} (\bibinfo {year} {2018})}\BibitemShut
  {NoStop}%
\bibitem [{\citenamefont {Lee}\ \emph {et~al.}(1957)\citenamefont {Lee},
  \citenamefont {Huang},\ and\ \citenamefont {Yang}}]{lee}%
  \BibitemOpen
  \bibfield  {author} {\bibinfo {author} {\bibfnamefont {T.~D.}\ \bibnamefont
  {Lee}}, \bibinfo {author} {\bibfnamefont {K.}~\bibnamefont {Huang}}, \ and\
  \bibinfo {author} {\bibfnamefont {C.~N.}\ \bibnamefont {Yang}},\ }\href@noop
  {} {\bibfield  {journal} {\bibinfo  {journal} {Physical Review}\ }\textbf
  {\bibinfo {volume} {106}},\ \bibinfo {pages} {1135} (\bibinfo {year}
  {1957})}\BibitemShut {NoStop}%
\bibitem [{\citenamefont {Petrov}(2015{\natexlab{a}})}]{petrov}%
  \BibitemOpen
  \bibfield  {author} {\bibinfo {author} {\bibfnamefont {D.}~\bibnamefont
  {Petrov}},\ }\href@noop {} {\bibfield  {journal} {\bibinfo  {journal}
  {Physical review letters}\ }\textbf {\bibinfo {volume} {115}},\ \bibinfo
  {pages} {155302} (\bibinfo {year} {2015}{\natexlab{a}})}\BibitemShut
  {NoStop}%
\bibitem [{\citenamefont {Petrov}\ and\ \citenamefont
  {Astrakharchik}(2016)}]{petrov_1d}%
  \BibitemOpen
  \bibfield  {author} {\bibinfo {author} {\bibfnamefont {D.}~\bibnamefont
  {Petrov}}\ and\ \bibinfo {author} {\bibfnamefont {G.}~\bibnamefont
  {Astrakharchik}},\ }\href@noop {} {\bibfield  {journal} {\bibinfo  {journal}
  {Physical review letters}\ }\textbf {\bibinfo {volume} {117}},\ \bibinfo
  {pages} {100401} (\bibinfo {year} {2016})}\BibitemShut {NoStop}%
\bibitem [{\citenamefont {Astrakharchik}\ and\ \citenamefont
  {Malomed}(2018)}]{malomed1}%
  \BibitemOpen
  \bibfield  {author} {\bibinfo {author} {\bibfnamefont {G.}~\bibnamefont
  {Astrakharchik}}\ and\ \bibinfo {author} {\bibfnamefont {B.~A.}\ \bibnamefont
  {Malomed}},\ }\href@noop {} {\bibfield  {journal} {\bibinfo  {journal}
  {Physical Review A}\ }\textbf {\bibinfo {volume} {98}},\ \bibinfo {pages}
  {013631} (\bibinfo {year} {2018})}\BibitemShut {NoStop}%
\bibitem [{\citenamefont {Debnath}\ and\ \citenamefont {Khan}(2021)}]{deb}%
  \BibitemOpen
  \bibfield  {author} {\bibinfo {author} {\bibfnamefont {A.}~\bibnamefont
  {Debnath}}\ and\ \bibinfo {author} {\bibfnamefont {A.}~\bibnamefont {Khan}},\
  }\href@noop {} {\bibfield  {journal} {\bibinfo  {journal} {Annalen der
  Physik}\ }\textbf {\bibinfo {volume} {533}},\ \bibinfo {pages} {2000549}
  (\bibinfo {year} {2021})}\BibitemShut {NoStop}%
\bibitem [{\citenamefont {Ferrier-Barbut}(2019)}]{ifb}%
  \BibitemOpen
  \bibfield  {author} {\bibinfo {author} {\bibfnamefont {I.}~\bibnamefont
  {Ferrier-Barbut}},\ }\href@noop {} {\bibfield  {journal} {\bibinfo  {journal}
  {Physics Today}\ }\textbf {\bibinfo {volume} {72}},\ \bibinfo {pages} {46}
  (\bibinfo {year} {2019})}\BibitemShut {NoStop}%
\bibitem [{\citenamefont {Tanzi}\ \emph {et~al.}(2019)\citenamefont {Tanzi},
  \citenamefont {Lucioni}, \citenamefont {Fam{\`a}}, \citenamefont {Catani},
  \citenamefont {Fioretti}, \citenamefont {Gabbanini}, \citenamefont {Bisset},
  \citenamefont {Santos},\ and\ \citenamefont
  {Modugno}}]{tanzi2019observation}%
  \BibitemOpen
  \bibfield  {author} {\bibinfo {author} {\bibfnamefont {L.}~\bibnamefont
  {Tanzi}}, \bibinfo {author} {\bibfnamefont {E.}~\bibnamefont {Lucioni}},
  \bibinfo {author} {\bibfnamefont {F.}~\bibnamefont {Fam{\`a}}}, \bibinfo
  {author} {\bibfnamefont {J.}~\bibnamefont {Catani}}, \bibinfo {author}
  {\bibfnamefont {A.}~\bibnamefont {Fioretti}}, \bibinfo {author}
  {\bibfnamefont {C.}~\bibnamefont {Gabbanini}}, \bibinfo {author}
  {\bibfnamefont {R.~N.}\ \bibnamefont {Bisset}}, \bibinfo {author}
  {\bibfnamefont {L.}~\bibnamefont {Santos}}, \ and\ \bibinfo {author}
  {\bibfnamefont {G.}~\bibnamefont {Modugno}},\ }\href@noop {} {\bibfield
  {journal} {\bibinfo  {journal} {Physical review letters}\ }\textbf {\bibinfo
  {volume} {122}},\ \bibinfo {pages} {130405} (\bibinfo {year}
  {2019})}\BibitemShut {NoStop}%
\bibitem [{\citenamefont {Chomaz}\ \emph {et~al.}(2019)\citenamefont {Chomaz},
  \citenamefont {Petter}, \citenamefont {Ilzh{\"o}fer}, \citenamefont {Natale},
  \citenamefont {Trautmann}, \citenamefont {Politi}, \citenamefont
  {Durastante}, \citenamefont {Van~Bijnen}, \citenamefont {Patscheider},
  \citenamefont {Sohmen} \emph {et~al.}}]{chomaz2019long}%
  \BibitemOpen
  \bibfield  {author} {\bibinfo {author} {\bibfnamefont {L.}~\bibnamefont
  {Chomaz}}, \bibinfo {author} {\bibfnamefont {D.}~\bibnamefont {Petter}},
  \bibinfo {author} {\bibfnamefont {P.}~\bibnamefont {Ilzh{\"o}fer}}, \bibinfo
  {author} {\bibfnamefont {G.}~\bibnamefont {Natale}}, \bibinfo {author}
  {\bibfnamefont {A.}~\bibnamefont {Trautmann}}, \bibinfo {author}
  {\bibfnamefont {C.}~\bibnamefont {Politi}}, \bibinfo {author} {\bibfnamefont
  {G.}~\bibnamefont {Durastante}}, \bibinfo {author} {\bibfnamefont
  {R.}~\bibnamefont {Van~Bijnen}}, \bibinfo {author} {\bibfnamefont
  {A.}~\bibnamefont {Patscheider}}, \bibinfo {author} {\bibfnamefont
  {M.}~\bibnamefont {Sohmen}},  \emph {et~al.},\ }\href@noop {} {\bibfield
  {journal} {\bibinfo  {journal} {Physical Review X}\ }\textbf {\bibinfo
  {volume} {9}},\ \bibinfo {pages} {021012} (\bibinfo {year}
  {2019})}\BibitemShut {NoStop}%
\bibitem [{\citenamefont {B{\"o}ttcher}\ \emph {et~al.}(2019)\citenamefont
  {B{\"o}ttcher}, \citenamefont {Schmidt}, \citenamefont {Wenzel},
  \citenamefont {Hertkorn}, \citenamefont {Guo}, \citenamefont {Langen},\ and\
  \citenamefont {Pfau}}]{bottcher2019transient}%
  \BibitemOpen
  \bibfield  {author} {\bibinfo {author} {\bibfnamefont {F.}~\bibnamefont
  {B{\"o}ttcher}}, \bibinfo {author} {\bibfnamefont {J.-N.}\ \bibnamefont
  {Schmidt}}, \bibinfo {author} {\bibfnamefont {M.}~\bibnamefont {Wenzel}},
  \bibinfo {author} {\bibfnamefont {J.}~\bibnamefont {Hertkorn}}, \bibinfo
  {author} {\bibfnamefont {M.}~\bibnamefont {Guo}}, \bibinfo {author}
  {\bibfnamefont {T.}~\bibnamefont {Langen}}, \ and\ \bibinfo {author}
  {\bibfnamefont {T.}~\bibnamefont {Pfau}},\ }\href@noop {} {\bibfield
  {journal} {\bibinfo  {journal} {Physical Review X}\ }\textbf {\bibinfo
  {volume} {9}},\ \bibinfo {pages} {011051} (\bibinfo {year}
  {2019})}\BibitemShut {NoStop}%
\bibitem [{\citenamefont {Chomaz}\ \emph {et~al.}(2018)\citenamefont {Chomaz},
  \citenamefont {van Bijnen}, \citenamefont {Petter}, \citenamefont {Faraoni},
  \citenamefont {Baier}, \citenamefont {Becher}, \citenamefont {Mark},
  \citenamefont {Waechtler}, \citenamefont {Santos},\ and\ \citenamefont
  {Ferlaino}}]{chomaz}%
  \BibitemOpen
  \bibfield  {author} {\bibinfo {author} {\bibfnamefont {L.}~\bibnamefont
  {Chomaz}}, \bibinfo {author} {\bibfnamefont {R.~M.}\ \bibnamefont {van
  Bijnen}}, \bibinfo {author} {\bibfnamefont {D.}~\bibnamefont {Petter}},
  \bibinfo {author} {\bibfnamefont {G.}~\bibnamefont {Faraoni}}, \bibinfo
  {author} {\bibfnamefont {S.}~\bibnamefont {Baier}}, \bibinfo {author}
  {\bibfnamefont {J.~H.}\ \bibnamefont {Becher}}, \bibinfo {author}
  {\bibfnamefont {M.~J.}\ \bibnamefont {Mark}}, \bibinfo {author}
  {\bibfnamefont {F.}~\bibnamefont {Waechtler}}, \bibinfo {author}
  {\bibfnamefont {L.}~\bibnamefont {Santos}}, \ and\ \bibinfo {author}
  {\bibfnamefont {F.}~\bibnamefont {Ferlaino}},\ }\href@noop {} {\bibfield
  {journal} {\bibinfo  {journal} {Nature physics}\ }\textbf {\bibinfo {volume}
  {14}},\ \bibinfo {pages} {442} (\bibinfo {year} {2018})}\BibitemShut
  {NoStop}%
\bibitem [{\citenamefont {Roccuzzo}\ and\ \citenamefont
  {Ancilotto}(2019)}]{roccuzzo2019supersolid}%
  \BibitemOpen
  \bibfield  {author} {\bibinfo {author} {\bibfnamefont {S.~M.}\ \bibnamefont
  {Roccuzzo}}\ and\ \bibinfo {author} {\bibfnamefont {F.}~\bibnamefont
  {Ancilotto}},\ }\href@noop {} {\bibfield  {journal} {\bibinfo  {journal}
  {Physical Review A}\ }\textbf {\bibinfo {volume} {99}},\ \bibinfo {pages}
  {041601} (\bibinfo {year} {2019})}\BibitemShut {NoStop}%
\bibitem [{\citenamefont {Pethick}\ and\ \citenamefont
  {Smith}(2008)}]{pethick2008bose}%
  \BibitemOpen
  \bibfield  {author} {\bibinfo {author} {\bibfnamefont {C.~J.}\ \bibnamefont
  {Pethick}}\ and\ \bibinfo {author} {\bibfnamefont {H.}~\bibnamefont
  {Smith}},\ }\href@noop {} {\emph {\bibinfo {title} {Bose--Einstein
  condensation in dilute gases}}}\ (\bibinfo  {publisher} {Cambridge university
  press},\ \bibinfo {year} {2008})\BibitemShut {NoStop}%
\bibitem [{\citenamefont {Pitaevskii}\ and\ \citenamefont
  {Stringari}(2016)}]{pitaevskii_2016bose}%
  \BibitemOpen
  \bibfield  {author} {\bibinfo {author} {\bibfnamefont {L.}~\bibnamefont
  {Pitaevskii}}\ and\ \bibinfo {author} {\bibfnamefont {S.}~\bibnamefont
  {Stringari}},\ }\href@noop {} {\emph {\bibinfo {title} {Bose-Einstein
  condensation and superfluidity}}},\ Vol.\ \bibinfo {volume} {164}\ (\bibinfo
  {publisher} {Oxford University Press},\ \bibinfo {year} {2016})\BibitemShut
  {NoStop}%
\bibitem [{\citenamefont {Mukherjee}\ \emph {et~al.}(2025)\citenamefont
  {Mukherjee}, \citenamefont {Saha},\ and\ \citenamefont
  {Dasgupta}}]{mukherjee2025collective}%
  \BibitemOpen
  \bibfield  {author} {\bibinfo {author} {\bibfnamefont {A.}~\bibnamefont
  {Mukherjee}}, \bibinfo {author} {\bibfnamefont {S.}~\bibnamefont {Saha}}, \
  and\ \bibinfo {author} {\bibfnamefont {R.}~\bibnamefont {Dasgupta}},\
  }\href@noop {} {\bibfield  {journal} {\bibinfo  {journal} {Journal of
  Physics: Condensed Matter}\ } (\bibinfo {year} {2025})}\BibitemShut {NoStop}%
\bibitem [{\citenamefont {Leggett}(1999)}]{leggett_1999superfluidity}%
  \BibitemOpen
  \bibfield  {author} {\bibinfo {author} {\bibfnamefont {A.~J.}\ \bibnamefont
  {Leggett}},\ }\href@noop {} {\bibfield  {journal} {\bibinfo  {journal}
  {Reviews of Modern Physics}\ }\textbf {\bibinfo {volume} {71}},\ \bibinfo
  {pages} {S318} (\bibinfo {year} {1999})}\BibitemShut {NoStop}%
\bibitem [{\citenamefont {Ferrier-Barbut}\ \emph {et~al.}(2018)\citenamefont
  {Ferrier-Barbut}, \citenamefont {Wenzel}, \citenamefont {B{\"o}ttcher},
  \citenamefont {Langen}, \citenamefont {Isoard}, \citenamefont {Stringari},\
  and\ \citenamefont {Pfau}}]{ferrier2018scissors}%
  \BibitemOpen
  \bibfield  {author} {\bibinfo {author} {\bibfnamefont {I.}~\bibnamefont
  {Ferrier-Barbut}}, \bibinfo {author} {\bibfnamefont {M.}~\bibnamefont
  {Wenzel}}, \bibinfo {author} {\bibfnamefont {F.}~\bibnamefont
  {B{\"o}ttcher}}, \bibinfo {author} {\bibfnamefont {T.}~\bibnamefont
  {Langen}}, \bibinfo {author} {\bibfnamefont {M.}~\bibnamefont {Isoard}},
  \bibinfo {author} {\bibfnamefont {S.}~\bibnamefont {Stringari}}, \ and\
  \bibinfo {author} {\bibfnamefont {T.}~\bibnamefont {Pfau}},\ }\href@noop {}
  {\bibfield  {journal} {\bibinfo  {journal} {Physical review letters}\
  }\textbf {\bibinfo {volume} {120}},\ \bibinfo {pages} {160402} (\bibinfo
  {year} {2018})}\BibitemShut {NoStop}%
\bibitem [{\citenamefont {Wysocki}\ \emph {et~al.}(2024)\citenamefont
  {Wysocki}, \citenamefont {Jachymski}, \citenamefont {Astrakharchik},\ and\
  \citenamefont {Tylutki}}]{P.wysocki_josephson_2024}%
  \BibitemOpen
  \bibfield  {author} {\bibinfo {author} {\bibfnamefont {P.}~\bibnamefont
  {Wysocki}}, \bibinfo {author} {\bibfnamefont {K.}~\bibnamefont {Jachymski}},
  \bibinfo {author} {\bibfnamefont {G.~E.}\ \bibnamefont {Astrakharchik}}, \
  and\ \bibinfo {author} {\bibfnamefont {M.}~\bibnamefont {Tylutki}},\
  }\href@noop {} {\bibfield  {journal} {\bibinfo  {journal} {Physical Review
  A}\ }\textbf {\bibinfo {volume} {110}},\ \bibinfo {pages} {033303} (\bibinfo
  {year} {2024})}\BibitemShut {NoStop}%
\bibitem [{\citenamefont {Abdullaev}\ \emph {et~al.}(2024)\citenamefont
  {Abdullaev}, \citenamefont {Galimzyanov},\ and\ \citenamefont
  {Shermakhmatov}}]{abdullaev2024beyond}%
  \BibitemOpen
  \bibfield  {author} {\bibinfo {author} {\bibfnamefont {F.~K.}\ \bibnamefont
  {Abdullaev}}, \bibinfo {author} {\bibfnamefont {R.~M.}\ \bibnamefont
  {Galimzyanov}}, \ and\ \bibinfo {author} {\bibfnamefont {A.~M.}\ \bibnamefont
  {Shermakhmatov}},\ }\href@noop {} {\bibfield  {journal} {\bibinfo  {journal}
  {The European Physical Journal D}\ }\textbf {\bibinfo {volume} {78}},\
  \bibinfo {pages} {118} (\bibinfo {year} {2024})}\BibitemShut {NoStop}%
\bibitem [{\citenamefont {Bardin}\ \emph {et~al.}(2024)\citenamefont {Bardin},
  \citenamefont {Lorenzi},\ and\ \citenamefont
  {Salasnich}}]{bardin2024quantum}%
  \BibitemOpen
  \bibfield  {author} {\bibinfo {author} {\bibfnamefont {A.}~\bibnamefont
  {Bardin}}, \bibinfo {author} {\bibfnamefont {F.}~\bibnamefont {Lorenzi}}, \
  and\ \bibinfo {author} {\bibfnamefont {L.}~\bibnamefont {Salasnich}},\
  }\href@noop {} {\bibfield  {journal} {\bibinfo  {journal} {New Journal of
  Physics}\ }\textbf {\bibinfo {volume} {26}},\ \bibinfo {pages} {013021}
  (\bibinfo {year} {2024})}\BibitemShut {NoStop}%
\bibitem [{\citenamefont {Bellettini}\ \emph {et~al.}(2024)\citenamefont
  {Bellettini}, \citenamefont {Richaud},\ and\ \citenamefont
  {Penna}}]{bellettini2024massive}%
  \BibitemOpen
  \bibfield  {author} {\bibinfo {author} {\bibfnamefont {A.}~\bibnamefont
  {Bellettini}}, \bibinfo {author} {\bibfnamefont {A.}~\bibnamefont {Richaud}},
  \ and\ \bibinfo {author} {\bibfnamefont {V.}~\bibnamefont {Penna}},\
  }\href@noop {} {\bibfield  {journal} {\bibinfo  {journal} {Physical Review
  Research}\ }\textbf {\bibinfo {volume} {6}},\ \bibinfo {pages} {043197}
  (\bibinfo {year} {2024})}\BibitemShut {NoStop}%
\bibitem [{\citenamefont {Biagioni}\ \emph {et~al.}(2024)\citenamefont
  {Biagioni}, \citenamefont {Antolini}, \citenamefont {Donelli}, \citenamefont
  {Pezz{\`e}}, \citenamefont {Smerzi}, \citenamefont {Fattori}, \citenamefont
  {Fioretti}, \citenamefont {Gabbanini}, \citenamefont {Inguscio},
  \citenamefont {Tanzi} \emph {et~al.}}]{biagioni2024measurement}%
  \BibitemOpen
  \bibfield  {author} {\bibinfo {author} {\bibfnamefont {G.}~\bibnamefont
  {Biagioni}}, \bibinfo {author} {\bibfnamefont {N.}~\bibnamefont {Antolini}},
  \bibinfo {author} {\bibfnamefont {B.}~\bibnamefont {Donelli}}, \bibinfo
  {author} {\bibfnamefont {L.}~\bibnamefont {Pezz{\`e}}}, \bibinfo {author}
  {\bibfnamefont {A.}~\bibnamefont {Smerzi}}, \bibinfo {author} {\bibfnamefont
  {M.}~\bibnamefont {Fattori}}, \bibinfo {author} {\bibfnamefont
  {A.}~\bibnamefont {Fioretti}}, \bibinfo {author} {\bibfnamefont
  {C.}~\bibnamefont {Gabbanini}}, \bibinfo {author} {\bibfnamefont
  {M.}~\bibnamefont {Inguscio}}, \bibinfo {author} {\bibfnamefont
  {L.}~\bibnamefont {Tanzi}},  \emph {et~al.},\ }\href@noop {} {\bibfield
  {journal} {\bibinfo  {journal} {Nature}\ }\textbf {\bibinfo {volume} {629}},\
  \bibinfo {pages} {773} (\bibinfo {year} {2024})}\BibitemShut {NoStop}%
\bibitem [{\citenamefont {Al-Marzoug}\ \emph {et~al.}(2026)\citenamefont
  {Al-Marzoug}, \citenamefont {Baizakov}, \citenamefont {Al~Khawaja},\ and\
  \citenamefont {Bahlouli}}]{al2026single}%
  \BibitemOpen
  \bibfield  {author} {\bibinfo {author} {\bibfnamefont {S.}~\bibnamefont
  {Al-Marzoug}}, \bibinfo {author} {\bibfnamefont {B.}~\bibnamefont
  {Baizakov}}, \bibinfo {author} {\bibfnamefont {U.}~\bibnamefont
  {Al~Khawaja}}, \ and\ \bibinfo {author} {\bibfnamefont {H.}~\bibnamefont
  {Bahlouli}},\ }\href@noop {} {\bibfield  {journal} {\bibinfo  {journal}
  {Chaos, Solitons \& Fractals}\ }\textbf {\bibinfo {volume} {206}},\ \bibinfo
  {pages} {117941} (\bibinfo {year} {2026})}\BibitemShut {NoStop}%
\bibitem [{\citenamefont {Khan}\ and\ \citenamefont {Debnath}(2022)}]{khan}%
  \BibitemOpen
  \bibfield  {author} {\bibinfo {author} {\bibfnamefont {A.}~\bibnamefont
  {Khan}}\ and\ \bibinfo {author} {\bibfnamefont {A.}~\bibnamefont {Debnath}},\
  }\href@noop {} {\bibfield  {journal} {\bibinfo  {journal} {Frontiers in
  Physics}\ }\textbf {\bibinfo {volume} {10}},\ \bibinfo {pages} {887338}
  (\bibinfo {year} {2022})}\BibitemShut {NoStop}%
\bibitem [{\citenamefont {Adusumalli}\ \emph {et~al.}(2024)\citenamefont
  {Adusumalli}, \citenamefont {Senapati}, \citenamefont {Singh},\ and\
  \citenamefont {Khan}}]{kathapla}%
  \BibitemOpen
  \bibfield  {author} {\bibinfo {author} {\bibfnamefont {S.~S.}\ \bibnamefont
  {Adusumalli}}, \bibinfo {author} {\bibfnamefont {K.}~\bibnamefont
  {Senapati}}, \bibinfo {author} {\bibfnamefont {S.}~\bibnamefont {Singh}}, \
  and\ \bibinfo {author} {\bibfnamefont {A.}~\bibnamefont {Khan}},\ }\href@noop
  {} {\bibfield  {journal} {\bibinfo  {journal} {Physics Letters A}\ }\textbf
  {\bibinfo {volume} {516}},\ \bibinfo {pages} {129638} (\bibinfo {year}
  {2024})}\BibitemShut {NoStop}%
\bibitem [{\citenamefont {Bulgac}(2002)}]{bulgac1}%
  \BibitemOpen
  \bibfield  {author} {\bibinfo {author} {\bibfnamefont {A.}~\bibnamefont
  {Bulgac}},\ }\href@noop {} {\bibfield  {journal} {\bibinfo  {journal}
  {Physical review letters}\ }\textbf {\bibinfo {volume} {89}},\ \bibinfo
  {pages} {050402} (\bibinfo {year} {2002})}\BibitemShut {NoStop}%
\bibitem [{\citenamefont {Alotaibi}\ \emph {et~al.}(2023)\citenamefont
  {Alotaibi}, \citenamefont {Al~Sakkaf},\ and\ \citenamefont
  {Al~Khawaja}}]{alotaibi2023unidirectional}%
  \BibitemOpen
  \bibfield  {author} {\bibinfo {author} {\bibfnamefont {M.}~\bibnamefont
  {Alotaibi}}, \bibinfo {author} {\bibfnamefont {L.}~\bibnamefont {Al~Sakkaf}},
  \ and\ \bibinfo {author} {\bibfnamefont {U.}~\bibnamefont {Al~Khawaja}},\
  }\href@noop {} {\bibfield  {journal} {\bibinfo  {journal} {Physics Letters
  A}\ }\textbf {\bibinfo {volume} {487}},\ \bibinfo {pages} {129120} (\bibinfo
  {year} {2023})}\BibitemShut {NoStop}%
\bibitem [{\citenamefont {Likharev}(1979)}]{likharev}%
  \BibitemOpen
  \bibfield  {author} {\bibinfo {author} {\bibfnamefont {K.}~\bibnamefont
  {Likharev}},\ }\href@noop {} {\bibfield  {journal} {\bibinfo  {journal}
  {Reviews of Modern Physics}\ }\textbf {\bibinfo {volume} {51}},\ \bibinfo
  {pages} {101} (\bibinfo {year} {1979})}\BibitemShut {NoStop}%
\bibitem [{\citenamefont {Anderson}\ \emph {et~al.}(1995)\citenamefont
  {Anderson}, \citenamefont {Ensher}, \citenamefont {Matthews}, \citenamefont
  {Wieman},\ and\ \citenamefont {Cornell}}]{anderson}%
  \BibitemOpen
  \bibfield  {author} {\bibinfo {author} {\bibfnamefont {M.~H.}\ \bibnamefont
  {Anderson}}, \bibinfo {author} {\bibfnamefont {J.~R.}\ \bibnamefont
  {Ensher}}, \bibinfo {author} {\bibfnamefont {M.~R.}\ \bibnamefont
  {Matthews}}, \bibinfo {author} {\bibfnamefont {C.~E.}\ \bibnamefont
  {Wieman}}, \ and\ \bibinfo {author} {\bibfnamefont {E.~A.}\ \bibnamefont
  {Cornell}},\ }\href@noop {} {\bibfield  {journal} {\bibinfo  {journal}
  {science}\ }\textbf {\bibinfo {volume} {269}},\ \bibinfo {pages} {198}
  (\bibinfo {year} {1995})}\BibitemShut {NoStop}%
\bibitem [{\citenamefont {Davis}\ \emph {et~al.}(1995)\citenamefont {Davis},
  \citenamefont {Mewes}, \citenamefont {Andrews}, \citenamefont {van Druten},
  \citenamefont {Durfee}, \citenamefont {Kurn},\ and\ \citenamefont
  {Ketterle}}]{davis}%
  \BibitemOpen
  \bibfield  {author} {\bibinfo {author} {\bibfnamefont {K.~B.}\ \bibnamefont
  {Davis}}, \bibinfo {author} {\bibfnamefont {M.-O.}\ \bibnamefont {Mewes}},
  \bibinfo {author} {\bibfnamefont {M.~R.}\ \bibnamefont {Andrews}}, \bibinfo
  {author} {\bibfnamefont {N.~J.}\ \bibnamefont {van Druten}}, \bibinfo
  {author} {\bibfnamefont {D.~S.}\ \bibnamefont {Durfee}}, \bibinfo {author}
  {\bibfnamefont {D.}~\bibnamefont {Kurn}}, \ and\ \bibinfo {author}
  {\bibfnamefont {W.}~\bibnamefont {Ketterle}},\ }\href@noop {} {\bibfield
  {journal} {\bibinfo  {journal} {Physical review letters}\ }\textbf {\bibinfo
  {volume} {75}},\ \bibinfo {pages} {3969} (\bibinfo {year}
  {1995})}\BibitemShut {NoStop}%
\bibitem [{\citenamefont {Debnath}\ \emph {et~al.}(2022)\citenamefont
  {Debnath}, \citenamefont {Tarun},\ and\ \citenamefont {Khan}}]{debnathjpb}%
  \BibitemOpen
  \bibfield  {author} {\bibinfo {author} {\bibfnamefont {A.}~\bibnamefont
  {Debnath}}, \bibinfo {author} {\bibfnamefont {J.}~\bibnamefont {Tarun}}, \
  and\ \bibinfo {author} {\bibfnamefont {A.}~\bibnamefont {Khan}},\ }\href@noop
  {} {\bibfield  {journal} {\bibinfo  {journal} {Journal of Physics B: Atomic,
  Molecular and Optical Physics}\ }\textbf {\bibinfo {volume} {55}},\ \bibinfo
  {pages} {025301} (\bibinfo {year} {2022})}\BibitemShut {NoStop}%
\bibitem [{\citenamefont {Raghavan}\ \emph {et~al.}(1999)\citenamefont
  {Raghavan}, \citenamefont {Smerzi}, \citenamefont {Fantoni},\ and\
  \citenamefont {Shenoy}}]{raghavan1999coherent}%
  \BibitemOpen
  \bibfield  {author} {\bibinfo {author} {\bibfnamefont {S.}~\bibnamefont
  {Raghavan}}, \bibinfo {author} {\bibfnamefont {A.}~\bibnamefont {Smerzi}},
  \bibinfo {author} {\bibfnamefont {S.}~\bibnamefont {Fantoni}}, \ and\
  \bibinfo {author} {\bibfnamefont {S.}~\bibnamefont {Shenoy}},\ }\href@noop {}
  {\bibfield  {journal} {\bibinfo  {journal} {Physical Review A}\ }\textbf
  {\bibinfo {volume} {59}},\ \bibinfo {pages} {620} (\bibinfo {year}
  {1999})}\BibitemShut {NoStop}%
\bibitem [{\citenamefont {Tononi}\ \emph {et~al.}(2019)\citenamefont {Tononi},
  \citenamefont {Wang},\ and\ \citenamefont {Salasnich}}]{tononi2019quantum}%
  \BibitemOpen
  \bibfield  {author} {\bibinfo {author} {\bibfnamefont {A.}~\bibnamefont
  {Tononi}}, \bibinfo {author} {\bibfnamefont {Y.}~\bibnamefont {Wang}}, \ and\
  \bibinfo {author} {\bibfnamefont {L.}~\bibnamefont {Salasnich}},\ }\href@noop
  {} {\bibfield  {journal} {\bibinfo  {journal} {Physical Review A}\ }\textbf
  {\bibinfo {volume} {99}},\ \bibinfo {pages} {063618} (\bibinfo {year}
  {2019})}\BibitemShut {NoStop}%
\bibitem [{\citenamefont {Strogatz}(2001)}]{strogatz2001nonlinear}%
  \BibitemOpen
  \bibfield  {author} {\bibinfo {author} {\bibfnamefont {S.~H.}\ \bibnamefont
  {Strogatz}},\ }\href@noop {} {\emph {\bibinfo {title} {Nonlinear dynamics and
  chaos: with applications to physics, biology, chemistry, and engineering
  (studies in nonlinearity)}}},\ Vol.~\bibinfo {volume} {1}\ (\bibinfo
  {publisher} {Westview press},\ \bibinfo {year} {2001})\BibitemShut {NoStop}%
\bibitem [{\citenamefont {Petrov}(2015{\natexlab{b}})}]{petrov1}%
  \BibitemOpen
  \bibfield  {author} {\bibinfo {author} {\bibfnamefont {D.~S.}\ \bibnamefont
  {Petrov}},\ }\href {\doibase 10.1103/PhysRevLett.115.155302} {\bibfield
  {journal} {\bibinfo  {journal} {Phys. Rev. Lett.}\ }\textbf {\bibinfo
  {volume} {115}},\ \bibinfo {pages} {155302} (\bibinfo {year}
  {2015}{\natexlab{b}})}\BibitemShut {NoStop}%
\bibitem [{\citenamefont {Paraoanu}\ \emph {et~al.}(2001)\citenamefont
  {Paraoanu}, \citenamefont {Kohler}, \citenamefont {Sols},\ and\ \citenamefont
  {Leggett}}]{paraoanu2001josephson}%
  \BibitemOpen
  \bibfield  {author} {\bibinfo {author} {\bibfnamefont {G.-S.}\ \bibnamefont
  {Paraoanu}}, \bibinfo {author} {\bibfnamefont {S.}~\bibnamefont {Kohler}},
  \bibinfo {author} {\bibfnamefont {F.}~\bibnamefont {Sols}}, \ and\ \bibinfo
  {author} {\bibfnamefont {A.}~\bibnamefont {Leggett}},\ }\href@noop {}
  {\bibfield  {journal} {\bibinfo  {journal} {Journal of Physics B: Atomic,
  Molecular and Optical Physics}\ }\textbf {\bibinfo {volume} {34}},\ \bibinfo
  {pages} {4689} (\bibinfo {year} {2001})}\BibitemShut {NoStop}%
\bibitem [{\citenamefont {Burchianti}\ \emph {et~al.}(2017)\citenamefont
  {Burchianti}, \citenamefont {Fort},\ and\ \citenamefont
  {Modugno}}]{burchianti2017josephson}%
  \BibitemOpen
  \bibfield  {author} {\bibinfo {author} {\bibfnamefont {A.}~\bibnamefont
  {Burchianti}}, \bibinfo {author} {\bibfnamefont {C.}~\bibnamefont {Fort}}, \
  and\ \bibinfo {author} {\bibfnamefont {M.}~\bibnamefont {Modugno}},\
  }\href@noop {} {\bibfield  {journal} {\bibinfo  {journal} {Physical Review
  A}\ }\textbf {\bibinfo {volume} {95}},\ \bibinfo {pages} {023627} (\bibinfo
  {year} {2017})}\BibitemShut {NoStop}%
\bibitem [{\citenamefont {Ala{\~n}a}\ \emph {et~al.}(2023)\citenamefont
  {Ala{\~n}a}, \citenamefont {Egusquiza},\ and\ \citenamefont
  {Modugno}}]{alana2023supersolid}%
  \BibitemOpen
  \bibfield  {author} {\bibinfo {author} {\bibfnamefont {A.}~\bibnamefont
  {Ala{\~n}a}}, \bibinfo {author} {\bibfnamefont {I.~L.}\ \bibnamefont
  {Egusquiza}}, \ and\ \bibinfo {author} {\bibfnamefont {M.}~\bibnamefont
  {Modugno}},\ }\href@noop {} {\bibfield  {journal} {\bibinfo  {journal}
  {Physical Review A}\ }\textbf {\bibinfo {volume} {108}},\ \bibinfo {pages}
  {033316} (\bibinfo {year} {2023})}\BibitemShut {NoStop}%
\bibitem [{\citenamefont {Singh}\ and\ \citenamefont
  {Khan}(2026)}]{singh2026fragmentation}%
  \BibitemOpen
  \bibfield  {author} {\bibinfo {author} {\bibfnamefont {S.}~\bibnamefont
  {Singh}}\ and\ \bibinfo {author} {\bibfnamefont {A.}~\bibnamefont {Khan}},\
  }\href@noop {} {\bibfield  {journal} {\bibinfo  {journal} {arXiv preprint
  arXiv:2609.20013}\ } (\bibinfo {year} {2026})}\BibitemShut {NoStop}%
\bibitem [{\citenamefont {Borysenko}\ \emph {et~al.}(2025)\citenamefont
  {Borysenko}, \citenamefont {Bazhan}, \citenamefont {Prykhodko}, \citenamefont
  {Pfeiffer}, \citenamefont {Lind}, \citenamefont {Birkl},\ and\ \citenamefont
  {Yakimenko}}]{borysenko2025acceleration}%
  \BibitemOpen
  \bibfield  {author} {\bibinfo {author} {\bibfnamefont {Y.}~\bibnamefont
  {Borysenko}}, \bibinfo {author} {\bibfnamefont {N.}~\bibnamefont {Bazhan}},
  \bibinfo {author} {\bibfnamefont {O.}~\bibnamefont {Prykhodko}}, \bibinfo
  {author} {\bibfnamefont {D.}~\bibnamefont {Pfeiffer}}, \bibinfo {author}
  {\bibfnamefont {L.}~\bibnamefont {Lind}}, \bibinfo {author} {\bibfnamefont
  {G.}~\bibnamefont {Birkl}}, \ and\ \bibinfo {author} {\bibfnamefont
  {A.}~\bibnamefont {Yakimenko}},\ }\href@noop {} {\bibfield  {journal}
  {\bibinfo  {journal} {Physical Review A}\ }\textbf {\bibinfo {volume}
  {111}},\ \bibinfo {pages} {043308} (\bibinfo {year} {2025})}\BibitemShut
  {NoStop}%
\end{thebibliography}%

\end{document}